\documentclass[12pt]{article}
\usepackage{latexsym,epsfig,graphicx,epstopdf,amsmath,amssymb,amscd,pifont, multirow,chicago,psfrag,paralist,dsfont}
\usepackage[titletoc]{appendix}
\usepackage{listings}
\usepackage{graphics}
\usepackage{float}
\usepackage{algorithm}
\usepackage[noend]{algpseudocode}
\usepackage{slashbox}

\usepackage{booktabs} % Required for better horizontal rules in tables
\usepackage{subcaption}
\usepackage{graphicx}
\usepackage{bm}

\usepackage{varwidth}
\usepackage{verbatim}

\usepackage[hyphens]{url}
\usepackage[hyperindex,breaklinks]{hyperref}
\usepackage[hyphenbreaks]{breakurl}

\newcommand{\loglik}{\mathcal{L}}

\newcommand{\Sigmavec}{{\boldsymbol{\Sigma}}}

\newcommand{\Ivec}{{\boldsymbol{I}}}
\newcommand{\lvec}{\boldsymbol{l}}
\newcommand{\betavec}{{\boldsymbol{\beta}}}

\newcommand{\EE}{\mathbb{E}}

\newcommand{\Xset}{\mathcal{X}}

\newcommand{\Cmat}{\mathbf{C}}

\newcommand{\Pmat}{\mathbf{P}}

\newcommand{\Gmat}{\mathbf{G}}
\newcommand{\zeromat}{\mathbf{0}}
\newcommand{\pvec}{\boldsymbol{p}}

\newcommand{\TAE}{{\textrm{TAE}}}
\newcommand{\MAE}{{\textrm{MAE}}}

\newcommand{\Var}{{\textrm{Var}}}

\newcommand{\ivec}{{\boldsymbol{i}}}

\newcommand{\N}{{\normalfont\textrm{N}}}
\newcommand{\diag}{\textrm{Diag}}
\newcommand{\PMD}{{\normalfont\textrm{PMD}}}
\newcommand{\wh}{\widehat}

\newcommand{\Xvec}{\boldsymbol{X}}
\newcommand{\Zvec}{\boldsymbol{Z}}
\newcommand{\xvec}{\boldsymbol{x}}
\newcommand{\gvec}{\boldsymbol{g}}

\newcommand{\muvec}{\boldsymbol{\mu}}

\newcommand{\code}{\texttt}

\newcommand{\Sig}{\boldsymbol{\Sigma}}
\newcommand{\mvec}{\boldsymbol{\mu}}

\newcommand{\SIM}{{\normalfont\textrm{SIM}}}
\newcommand{\NA}{{\textrm{NA}}}
\newcommand{\dft}{{\textrm{DFT-CF}}}

\newtheorem{example}{Example}
\newtheorem{thm}{Theorem}
\newtheorem{ppt}{Proposition}

\begin{document}
%%%%%%%%%%%%TITLE%%%%%%%%%%%%%%%%%%%%%%%%%%%%%%%%%%%

\title{The Poisson Multinomial Distribution and Its Applications in Voting Theory, Ecological Inference, and Machine Learning}

%\iffalse
\author{
Zhengzhi Lin, Yueyao Wang, and Yili Hong\\[1.5ex]
{Department of Statistics, Virginia Tech, Blacksburg, VA 24061}
}
%\fi
	
%\date{\today}
\date{}

\maketitle
	%%%%%%%%%%%%%%%%%%%%%%%%%%%%%%%%%%%%%%%%%%%%%%%%%%%%%%%%%%%%%%%%%%%%%%%%%%%%%%%%%%%%%%%%%%%%%%%%%%%%%%%%%%%%%%%%
\begin{abstract}
The Poisson multinomial distribution (PMD) describes the distribution of the sum of $n$ independent but non-identically distributed random vectors, in which each random vector is of length $m$ with 0/1 valued elements and only one of its elements can take value~1 with a certain probability. Those probabilities are different for the $m$ elements across the $n$ random vectors, and form an $n \times m$ matrix with row sum equals to 1. We call this $n\times m$ matrix the success probability matrix (SPM). Each SPM uniquely defines a $\PMD$. The $\PMD$ is useful in many areas such as, voting theory, ecological inference, and machine learning. The distribution functions of $\PMD$, however, are usually difficult to compute. In this paper, we develop efficient methods to compute the probability mass function (pmf) for the PMD using multivariate Fourier transform, normal approximation, and simulations. We study the accuracy and efficiency of those methods and give recommendations for which methods to use under various scenarios. We also illustrate the use of the $\PMD$ via three applications, namely, in voting probability calculation, aggregated data inference, and uncertainty quantification in classification. We build an R package that implements the proposed methods, and illustrate the package with examples.

\textbf{Key Words:} Aggregated Data Inference; Classification; Multinomial Distribution; Poisson Binomial Distribution; Political Science; Uncertainty Quantification.
\end{abstract}
	
	%%%%%%%%%%%%%%%%%%%%%%%%%%%%%%%%%%%%%%%%%%%%%%%%%%%%%%%%%%%%%%%%%%%%%%%%%%%%%%%%%%%%%%%%%%%%%%%%%%%%%%%%%%%%%%%%%%%%%
%\newpage
%\tableofcontents
\newpage
	%%%%%%%%%%%%%%%%%%%%%%%%%%%%%%%%%%%%%%%%%%%%%%%%%%%%%%%%%%%%%%%%%%%%%%%%%%%%%%%%%%%%%%%%%%%%%%%%%%%%%%%%%%%%%%%%%%%%%
\section{Introduction}
%%%%%%%%%%%%%%%%%%%%%%%%%%%%%%%%%%%%%%%%%%%%%%%%%%%%%%%%%%%%%%%%%%%%%%%%%%%%%%%%%%%%%%%%%%%%%%%%%%%%%%%%%%%%%%%%%%%%%
\subsection{Motivation}
%%%%%%%%%%%%%%%%%%%%%%%%%%%%%%%%%%%%%%%%%%%%%%%%%%%%%%%%%%%%%%%%%%%%%%%%%%%%%%%%%%%%%%%%%%%%%%%%%%%%%%%%%%%%%%%%%%%%%

Suppose there are $n$ independent but non-identically distributed random vectors. Each of the vectors is of length $m$ with 0/1 valued elements. For each vector, only one of its elements can take value 1 with a certain success probability. Those success probabilities vary from elements across those $n$ random vectors. The Poisson multinomial distribution (PMD) describes the distribution of the sum of those $n$ random vectors. Those success probabilities can be arranged into an $n \times m$ matrix, which we call the success probability matrix (SPM), and the sum of each row of the SPM is 1. Each SPM uniquely defines a $\PMD$.

For a simple example, suppose there are $n$ balls, and one needs to throw them into $m$ different bins. For each ball, it will fall into one of those $m$ bins, but the probabilities are different from bin to bin. In addition, each ball has its own probability of falling into a specific bin. Then the probability distribution of the ball counts in each bin is the $\PMD$. The $\PMD$ is a generalization of multinomial distribution in which the probabilities are identical from ball to ball. In addition, when $m=2$, the $\PMD$ reduces to the Poisson binomial distribution (e.g., \citeNP{Hong2013}). The $\PMD$ has applications in many fields, including voting theory in political science, ecological inference, and uncertainty quantification in machine learning. We give some detailed examples below.

In an election scenario, suppose a committee with $n$ members needs to elect a chairman from $m$ candidates. Each member has different voting behavior so that the probabilities of voting for each candidate are different. An election outcome is the number of votes that each candidate receives after the voting. The $\PMD$ can be used to describe the distribution of the vote counts. Questions like the following are often asked. What is the most likely election result? What is the probability that a specific candidate wins the election? These questions are often of interest but hard to answer, because there are many possible election outcomes (i.e., $n+m-1$ choose $m-1$, denoted by $\binom{n+m-1}{m-1}$) based on the votes that each candidate receives. However, if we can compute the probability mass function (pmf) of the $\PMD$, those questions for the voting outcomes can be answered.

In ecological inference, the goal is to draw conclusions about individual-level behavior using aggregate-level data (e.g., \citeNP{Schuessler1999}). Statistical models are often used in making inference from aggregated data. An aggregated dataset is obtained by combining raw data into groups. The aggregated data may contain no or partial individual information. Instead, the data contain summary information of groups (e.g., the counts for each group). For example, we consider an application with a categorical response variable. The raw data are separated into groups. Then in each group, the counts of each category are available for analysis, which can be described by a random vector that follows a $\PMD$. If one can calculate the likelihood based on the $\PMD$, then the statistical inference can be carried out based on the aggregated data.

In machine learning, people often encounter the need to classify observations into different categories. Suppose a machine-learning model computes the probabilities of an observation falling into one of the $m$ categories. Although the decision is usually made by selecting the category that has the highest probability, in some scenarios, the category is assigned according to the computed probability for each category. That is, the decision is made by drawing a random sample from a one-hot random vector with respect to the computed probabilities. If we have $n$ observations, there will be $n$ independent categorical distributions. In the uncertainty quantification context, one can consider the counts that the classifier put into each category and form a confusion matrix. The $\PMD$ can be used to characterize the probability distribution of the counts in the confusion matrix. Thus, the PMD is useful if one is interested in understanding the uncertainty in the confusion matrix.

From the above discussions, we can see that $\PMD$ has potential applications in many areas. However, the computing of its pmf is non-trivial, and there is no efficient algorithm available to compute the pmf. In theory, enumeration can be used because one can list all possible outcomes.  However, enumeration is only feasible when both $n$ and $m$ are small. When $n$ and $m$ increase, the computing by enumeration will be impractical. There is a need for methods that are computationally efficient.  Therefore, we are motivated to develop methods that can compute the pmf of $\PMD$ efficiently.

%%%%%%%%%%%%%%%%%%%%%%%%%%%%%%%%%%%%%%%%%%%%%%%%%%%%%%%%%%%%%%%%%%%%%%%%%%%%%%%%%%%%%%%
\subsection{Related Literature and Contribution of This Work}

Some previous work studied the structure of the $\PMD$ and some of its properties. \citeN{Daskalakis2015OnTS} proved that $\PMD$ is $\epsilon$-cover, which means there exists a set of distributions small enough to cover the set of all $\PMD$.  \citeN{diakonikolas2016fourier} obtained a different understanding of the structure of $\PMD$ using Fourier transform and disclosed the sparsity of $\PMD$. On related topics, \citeN{Hong2013} considered both exact and approximate methods for computing the pmf of Poisson binomial distribution, which is a special case of the PMD. \citeN{zhang2018generalized} introduced the generalized Poisson binomial distribution and developed an algorithm to compute its distribution functions. \citeN{BiscarriZhaoBrunner2018} developed a convolution scheme and improved the computational efficiency of the pmf of Poisson binomial distribution. \citeN{pbd} built an R package that includes methods developed in \citeN{Hong2013}, \citeN{zhang2018generalized}, and \citeN{BiscarriZhaoBrunner2018}. So far, there are no efficient algorithms and software implementation for computing the pmf of PMD.

Current studies illustrated that the $\PMD$ can be applied to many areas. In game theory, based on the structure and properties of the $\PMD$, \citeN{diakonikolas2016fourier} constructed an efficient scheme to approximate the Nash equilibrium in anonymous games. \citeN{Cheng2017PlayingAG} proved that any $n$-player anonymous games can have an approximate Nash equilibrium in polynomial time. \citeN{akter2019double} introduced $\PMD$ in the image processing field for the first time, and used $\PMD$ to process the encrypted image data and computed the pmf of $\PMD$ to obtain optimal results. However, the pmf of $\PMD$ was computed by using simplified multinomial distributions, which is an approximation.

The contributions of this paper are as follows. We develop an exact method to compute the pmf of PMD that uses the discrete Fourier transform (DFT) of the characteristic function (CF).  The method is called the $\dft$ method. We also construct two approximation methods to compute the pmf of $\PMD$, which are based on normal approximations and simulations. We study the accuracy of each method under various scenarios and explore the time efficiency of the $\dft$ method. We illustrate the applications of $\PMD$ in the context of voting theory in political science, ecological inference, and uncertainty quantification in machine learning. We also build an R package that includes the developed methods, and demonstrate the use of the package.

\subsection{Overview}
The rest of the paper is organized as follows. In Section~\ref{sec:pmd}, we describe the formal definition of the PMD and discuss its properties. In Section~\ref{sec:cmpt.pmf}, we develop three methods to compute the pmf of $\PMD$ and provide some theoretical results that provide insights on the error bounds of approximation methods. In Section~\ref{sec:Method Comparisons}, we study the accuracy and time efficiency of the three methods, and provide recommendations on the use of each method. In Section~\ref{sec:applications}, we illustrate our methods via three applications in voting theory, ecological inference, and machine learning. In Section~\ref{sec:rpackage}, we develop an R package for the PMD and illustrate its use. In Section~\ref{sec:conclusion}, we conclude the paper and describe some areas for future research.

%%%%%%%%%%%%%%%%%%%%%%%%%%%%%%%%%%%%%%%%%%%%%%%
\section{Poisson Multinomial Distribution} \label{sec:pmd}
%%%%%%%%%%%%%%%%%%%%%%%%%%%%%%%%%%%%%%%%%%%%%%%%%%%%%%%%%%%%%%%%%%%%%%%%%%%%%%%%%%
\subsection{Definition of the Distribution} \label{subsec:def}
Let $\Ivec_{i} = (I_{i1}, \dots, I_{im})', i=1, \dots, n$ be independent vectors of random indicators (i.e., 0/1 valued). Here, $n$ is the number of random vectors and $m$ is the number of categories. Let the associated probabilities be $\pvec_{i} = (p_{i1}, \dots, p_{im})'$. That is, $\Pr(I_{ij}=1)=p_{ij}$, $i=1,\ldots, n,$ and  $j=1, \ldots, m$. Note that, for a given $i$, we have $\sum_{j=1}^{m}p_{ij}=1$ and $\sum_{j=1}^{m}I_{ij}=1$.

The sum of the random vectors, $\Xvec = (X_{1}, \dots, X_{m})'= \sum_{i=1}^{n}\Ivec_{i}$, follows a PMD. We denote it as,
$$\Xvec \sim \PMD(\Pmat).$$ where $X_{j} = \sum_{i=1}^{n} I_{ij}, j=1,\dots,m$. Here, the $n \times m$ matrix $\Pmat$ is called the success probability matrix (SPM), which is,
\begin{align*}
\Pmat = (\pvec_{1}, \dots, \pvec_{n} )'.
\end{align*}
Note that the random variables, $X_1, \dots, X_{m}$, satisfy the constraint that $\sum_{j=1}^{m}X_{j} = n.$ Hence, we can replace one of the elements in $\Xvec$, for example, $X_m$ with $n-\sum_{j=1}^{m-1}X_j$.

For some special cases, when the SPM is identical across all rows, that is the $\Ivec_{i}$'s are identically distributed, the distribution of $\Xvec$ reduces to the multinomial distribution. Hence, the $\PMD$ is a generalization of the multinomial distribution. When $m=2$, the $\PMD$ reduces to the Poisson binomial distribution, which can be further reduced to the binomial distribution.

Let $\xvec = (x_1,\dots,x_m)'$. The support set of the PMD is,
\begin{align*}
\Xset=\left\{\xvec\in \mathbb{N}^m: \sum_{j=1}^{m}x_j=n\right\},
\end{align*}
where $\mathbb{N}$ is the set of non-negative integers. Note that $\Pr(\Xvec=\xvec)>0$ for $\xvec\in\Xset$, and the number of elements in $\Xset$ is $h(n,m)=\binom{n+m-1}{m-1}$.

The probability mass function (pmf) of PMD is defined as,
$$p(\xvec)=\Pr(\Xvec=\xvec) = \text{Pr} \left( X_1 = x_1, \dots, X_m = x_{m-1}, X_{m} = n-\sum_{i=1}^{m}x_i \right),$$
for $\xvec\in\Xset$. Here we give a simple example in elections to illustrate the calculation of the pmf by enumeration.

\begin{example}\label{eg:simple.example}
Suppose there are four voters and three candidates in an election. That is, $n=4$ and $m=3$.   The resulting vote counts can be modeled by a random vector $\Xvec \sim \PMD(\Pmat)$. Based on historical information, the SPM is obtained as,
\begin{align*}
\Pmat_{4 \times 3} =
\begin{pmatrix}
0.1 &  0.2 & 0.7\\
0.5 & 0.2 & 0.3\\
0.4 &  0.5 & 0.1\\
0.8 & 0.1 & 0.1
\end{pmatrix}.
\end{align*}
The number of distinct outcomes of the election is $h(n,m)=15$ in this case. The pmf of $\Xvec$ can be computed by enumeration. For example, let $\xvec = (4,0,0)'$. The probability of the outcome that the first candidate receives 4 votes and others receives 0 vote is $\Pr(\Xvec=\xvec)$, which is,
\begin{align*}
\Pr\left( \Xvec = \xvec \right) = 0.1\times 0.5 \times 0.4 \times 0.8 = 0.016.
\end{align*}
For another example, if we take $\xvec=(1,3,0)'$, the probability of $\Xvec = \xvec$ is,
\begin{align*}
\Pr\left( \Xvec = \xvec \right)  =  & 0.1\times 0.2 \times 0.5 \times 0.1 +
 0.5\times0.2\times0.5 \times 0.1 \\
 & + 0.4\times0.2\times0.2\times0.1 + 0.8\times0.2\times0.2\times0.5 = 0.0236.
\end{align*}

\end{example}

In this example, both $n$ and $m$ are small so that enumeration can be used to calculate the exact probability. It becomes impractical to enumerate all possible outcomes because $h(n, m)$ explodes as $n$ and $m$ increase, even for moderate values of $n$ and $m$. For example, when $n=20$ and $m=5$, there will be $h(20, 5)=10626$ possible outcomes to enumerate.

%%%%%%%%%%%%%%%%%%%%%%%%%%%%%%%%%%%%%%%%%%%%%%%%%%%%%%%%%%%%%%%%%%%%%%%%%%%%%%%%%
\subsection{Properties of the Distribution}\label{subsec:propty}
In this section, we present some results on the properties of the $\PMD$.

\begin{ppt}%\normalfont
Suppose $\Xvec \sim \PMD (\Pmat)$. The mean of $\Xvec$ is,
$$\EE(\Xvec) = \muvec = \left( p_{\cdot1},\dots,p_{\cdot m}\right)',$$
where $p_{\cdot j} = \sum_{i=1}^{n}p_{ij}$.
The variance-covariance matrix of $\Xvec$, denoted by $\Sig$, is an $m \times m$ matrix with entry $\sigma_{jk}$ calculated as,
\begin{align*}
   \sigma_{jk} =
           \begin{cases}
             \sum_{i=1}^{n}p_{ij}(1-p_{ij}) & \text{ if } j=k\\
             -\sum_{i=1}^{n}p_{ij}p_{ik} &  \text{ if } j \neq k\\
           \end{cases}.
\end{align*}
The CF for the $\PMD$ is,
\begin{align*}
\phi(t_1, \dots, t_{m})  =  \sum_{x_1 = 0}^{n} \cdots \sum_{x_{m} = 0}^n p(\xvec)\exp\left(\ivec\sum_{j=1}^{m}t_jx_j\right).
\end{align*}
where $\xvec=(x_1, \dots, x_j, \dots, x_m)'$ and $\ivec=\sqrt{-1}$.
\end{ppt}

The derivations of the mean and CF are straightforward by following the corresponding definitions. The $\Sig$ can be calculated by noting that for any fixed $i=1,\dots,n$, $I_{ij}$ and $I_{ik}$ has covariance $-p_{ij}p_{ik},j=1,\dots,m,k=1,\dots,m$. Note that the covariance matrix $\Sig$ is singular, because the elements of $\Xvec$ are linearly dependent.

For notation convenience and later development of computational algorithms, we introduce a reduced version of $\Xvec$, denoted by $\Xvec^{\ast}$. Recall that $\sum_{j=1}^{m}X_j = n$. Without loss of generality, we drop the last element of $\Xvec$. Let $\Ivec_i^{\ast} = (I_{i1},\dots,I_{i,m-1})'$, and we denote,
$$\Xvec^{\ast}=\sum_{i=1}^{n}\Ivec_{i}^{\ast}=(X_1,\dots,X_{m-1})',$$
with corresponding $\xvec^{\ast} = (x_1,\dots,x_{m-1})'$ and $\Pmat^{\ast}$ equals to the first $m-1$ columns of $\Pmat$. That is,
$$
\Pmat^{\ast} = \left(\pvec_1^{\ast},\dots, \pvec_n^{\ast} \right)',
$$
where $\pvec_{i}^{\ast} = \left(p_{i1},\dots,p_{i,m-1} \right)', i = 1,\dots,n$. It is easy to see that $\Xvec$ and $\Xvec^{\ast}$ uniquely determine each other.

We denote the mean of $\Xvec^{\ast}$ as $\muvec^{\ast}$. That is,
\begin{align}\label{eq:X.star.mu}
\EE(\Xvec^{\ast}) =\muvec^{\ast} = \left( p_{\cdot1},\dots,p_{\cdot,m-1}\right)'.
\end{align}
Also, we denote the $(m-1) \times (m-1)$ covariance matrix of $\Xvec^{\ast}$ as $\Sigmavec^{\ast}$, which is non-singular. We have,
\begin{align}\label{eq:X.star.Sigma}
\Var(\Xvec^{\ast}) =\Sig^{\ast}=\sum_{i=1}^n[\diag(\pvec_i^{\ast})-\pvec_i^{\ast} \pvec_i^{\ast \prime}],
\end{align}
where $\pvec_i^{\ast}$ is the $i$th row of $\Pmat^{\ast}$, and $\diag(\pvec_i^{\ast})$ is a diagonal matrix with the main diagonal elements being $	\pvec_i^{\ast}$.
In addition, the CF of $\Xvec^{\ast}$ is,
\begin{align*}
\phi^{\ast}(t_1, \dots, t_{m-1})  =  \sum_{x_1 = 0}^{n} \cdots \sum_{x_{m-1} = 0}^n p^{\ast}(\xvec^{\ast})\exp\left(\ivec\sum_{j=1}^{m-1}t_jx_j\right),
\end{align*}
where $\xvec^{\ast}=(x_1,\ldots,x_{j},\ldots, x_{m-1})$, and
\begin{align}\label{eqn:p.star.x.star}
p^{\ast}(\xvec^{\ast})=\Pr(\Xvec^{\ast}=\xvec^{\ast})=p(\xvec).
\end{align}

When the SPM $\Pmat$ is a block-diagonal matrix, the calculation of the pmf can be greatly simplified, which is given as follows.
\begin{ppt}\label{prop:Pmat.product}
Suppose $\Xvec \sim \PMD(\Pmat)$. If the SPM $\Pmat$ can be written as a block-diagonal matrix as,
$$
\Pmat=
\begin{pmatrix}
  \Pmat_1  & \zeromat & \cdots & \zeromat \\
  \zeromat & \Pmat_2  & \cdots & \zeromat \\
  \vdots   & \vdots   & \ddots & \vdots   \\
  \zeromat & \zeromat & \cdots & \Pmat_s  \\
\end{pmatrix},
$$
where $s$ is the number of blocks. In addition, consider independent random vectors $\Xvec_{k} \sim \PMD(\Pmat_{k}), k=1,\dots,s$. Let $\xvec$ be a point from the support set of $\Xvec$. We partition $\xvec$ according to the dimensions of $\Pmat_k$'s as,
$$\xvec= (\xvec_{1}',\dots,\xvec_{s}')'.$$
The pmf of $\Xvec$ can be calculated as the product of the corresponding marginal pmfs of those $\Xvec_k$'s. That is,
\begin{align*}
p(\xvec) = \Pr(\Xvec=\xvec)= \Pr(\Xvec_{1}=\xvec_{1}, \dots, \Xvec_{s}=\xvec_{s})= \prod_{k=1}^s \Pr(\Xvec_{k} = \xvec_{k}).
\end{align*}

\end{ppt}

To show Proposition~\ref{prop:Pmat.product} in a heuristic way, we denote the size of $\Pmat_{k}$ as $n_k \times m_k$, and note that $\sum_{k=1}^s n_k = n$ and $\sum_{k=1}^s m_k = m$. Suppose there are $n$ voters for $m$ candidates, a certain group of voters only vote for a certain candidate and there are no overlaps. Thus, we can separate candidates and voters into independent groups. In group $k$, $k = 1,\dots,s$, voters voting for the corresponding candidate according to SPM $\Pmat_{k}$. Thus, the probability of the overall voting result can be calculated by the product of the probability of the voting result from each group. A rigorous proof can be done by the decomposition of the CF of $\Xvec$ into the CFs of those $\Xvec_k$'s.

%%%%%%%%%%%%%%%%%%%%%%%%%%%%%%%%%%%%%%%%%%%%%%%%%%%%%%%%%%%%%%%%%%%%%%%%%%%%%%%%%%%%
\section{Computation of The Probability Mass Function} \label{sec:cmpt.pmf}
%%%%%%%%%%%%%%%%%%%%%%%%%%%%%%%%%%%%%%%%%%%%%%%%%%%%%%%%%%%%%%%%%%%%%%%%%%%%%%%%%%%%
We introduce three methods for computing the pmf, which are the method based on multi-dimensional DFT of the CF of the PMD (denoted as $\dft$), the normal approximation method (denoted as $\NA$), and the simulation-based method (denoted as $\SIM$). The $\dft$ method is an exact method, while the other two are approximate methods.

%%%%%%%%%%%%%%%%%%%%%%%%%%%%%%%%%%%%%%%%%%%%%%%%%%%%%%%%%%%%%%%%%%%%%%%%%%%%%%%%%%%%%%%%%%%%%%%%%%
\subsection{The $\dft$ Method}
%%%%%%%%%%%%%%%%%%%%%%%%%%%%%%%%%%%%%%%%%%%%%%%%%%%%%%%%%%%%%%%%%%%%%%%%%%%%%%%%%%%%%%%%%%%%%%%%%%
In this section, we describe the $\dft$ method. Although there is no closed-form expression for the pmf of PMD, the CF of the distribution can be calculated explicitly. The CF of the reduced version, $\Xvec^{\ast}$, is
\begin{align*}
\phi^{\ast}(t_1, \dots, t_{m-1}) & = \EE\left[\exp\left(\ivec\sum_{j=1}^{m-1}t_jX_j\right)\right]=\EE\left[\exp\left(\ivec\sum_{i = 1}^n \sum_{j=1}^{m-1}t_j I_{ij}\right)\right].
\end{align*}
Here $\ivec=\sqrt{-1}$. By the definition of CF, we have,
\begin{align}\label{eq:2}
\EE\left[\exp\left(\ivec\sum_{j=1}^{m-1}t_jX_j\right)\right] = \sum_{x_1 = 0}^{n}\cdots \sum_{x_{m-1} = 0}^n p^{\ast}(\xvec^{\ast})\exp\left(\ivec\sum_{j=1}^{m-1}t_jx_j\right),
\end{align}
where $\xvec^{\ast}$ and $p^{\ast}(\xvec^{\ast})$ are defined in \eqref{eqn:p.star.x.star}. By the definition of the PMD,
\begin{align}\label{eq:3}
\exp\left(\ivec\sum_{j=1}^{m-1}t_jX_j\right)= \exp\left(\ivec\sum_{i = 1}^n \sum_{j=1}^{m-1}t_j I_{ij}\right).
\end{align}
The expectation of the right-hand side of (\ref{eq:3}) can be expressed as,
\begin{align}\label{eq:4}
\EE\left[\exp\left(\ivec\sum_{i = 1}^n \sum_{j=1}^{m-1}t_j I_{ij}\right)\right]& = \EE\left[ \exp\left( \ivec\sum_{j=1}^{m-1} t_jI_{1j} + \dots + \ivec\sum_{j=1}^{m-1} t_jI_{nj}\right)\right]\\\nonumber
& = \prod_{i=1}^n \EE\left[ \exp\left( \ivec \sum_{j=1}^{m-1} t_j I_{ij}\right)\right] = \prod_{i=1}^n \left[p_{im}+\sum_{j=1}^{m-1}p_{ij}\exp(\ivec t_j)\right].
\end{align}
We know \eqref{eq:4} equals to (\ref{eq:2}). Therefore, we obtain,
\begin{align}\label{eq:5}
\sum_{x_1 = 0}^{n}\cdots \sum_{x_{m-1} = 0}^n p^{\ast}(\xvec^{\ast})\exp\left(\ivec\sum_{j=1}^{m-1}t_jx_j\right)= \prod_{i=1}^{n}\left[p_{im}+\sum_{j=1}^{m-1}p_{ij}\exp(\ivec t_j)\right].
\end{align}
Let $t_j = \omega l_j$, $l_j = 0, \ldots, n$, $\omega = 2\pi/(n+1)$. Then \eqref{eq:5} becomes,
\begin{align}\label{eq:6}
\frac{1}{(n+1)^{m-1}} \sum_{x_1 = 0}^{n}\cdots \sum_{x_{m-1} = 0}^n p^{\ast}(\xvec^{\ast}) \exp\left(\ivec\omega\sum_{j=1}^{m-1}l_j x_j\right)= \frac{q(\lvec)}{(n+1)^{m-1}},
\end{align}
where $\lvec = (l_1,\dots,l_{m-1})'$, and
$$
q(\lvec)=\prod_{i=1}^{n}\left[p_{im}+\sum_{j=1}^{m-1}p_{ij}\exp(\ivec \omega l_j)\right].
$$	
Note that $q(\lvec)$ can be computed directly. Notice the left-hand side of \eqref{eq:6} is the inverse multi-dimensional DFT of the multi-dimensional array $\{p^{\ast}(\xvec^{\ast}), x_i = 0, \dots, n; i=1, \ldots, (m-1)\}$, which is of dimension, $$\underbrace{(n+1)\times\cdots\times(n+1)}_{(m-1)\textrm{ times }}.$$
Therefore, we can apply multi-dimensional DFT on both sides to recover the multi-dimensional array. The pmf can be obtained as,
\begin{align}\label{eq:dft.cf.formula}
p^{\ast}(\xvec^{\ast}) = \frac{1}{(n+1)^{m-1}}\sum_{l_1 = 0}^{n}\cdots \sum_{l_{m-1} = 0}^n q(\lvec) \exp\left(-\ivec\omega\sum_{j=1}^{m-1}l_j x_j\right).
\end{align}

Note that we have $(n+1)^{m-1}$ different $\lvec$, as $l_j$ varies from $0$ to $n$ and $j$ varies from 1 to $(m-1)$. For example, if we have $n=4$, $m=4$, then $\lvec$ can be $(0, 0, 0)', (0, 0, 1)', \dots, (4, 4, 4)'$, and we have 125 different vectors in total. For each $\lvec$, we can compute the corresponding $q(\lvec)$. Then we can use \eqref{eq:dft.cf.formula} to compute the pmf $p(\xvec)=p^{\ast}(\xvec^{\ast})$. To speed up the computing of \eqref{eq:dft.cf.formula}, we apply the fast Fourier transform (FFT) algorithm. The FFT algorithm is available in libraries such as the FFTW3 (\citeNP{FrigoJohnson2005}).

%%%%%%%%%%%%%%%%%%%%%%%%%%%%%%%%%%%%%%%%%%%%%%%%%%%%%%%%%%%%%%%%%%%%%%%%%%%%%%%%%%%%%%%%%%%%%%%%%
\subsection{Normal-Approximation Based Method}
%%%%%%%%%%%%%%%%%%%%%%%%%%%%%%%%%%%%%%%%%%%%%%%%%%%%%%%%%%%%%%%%%%%%%%%%%%%%%%%%%%%%%%%%%%%%%%%%%

The normal approximation ($\NA$) method uses the central limit theorem (CLT) to approximate the distribution of the $\PMD$.  Because the covariance matrix of $\Xvec$ is singular, we work with the reduced version $\Xvec^{\ast}$ to establish the normal approximation result. Recall that $\Xvec^{\ast}$ has mean $\muvec^{\ast}$ and variance-covariance matrix $\Sigmavec^{\ast}$, which are defined in \eqref{eq:X.star.mu} and \eqref{eq:X.star.Sigma}, respectively.

By the CLT (\citeNP{Daskalakis2015OnTS}), $\Xvec^{\ast}$ is asymptotically distributed with $\N(\muvec^{\ast}, \Sig^{\ast})$. That is,
$$\Xvec^{\ast} \dot\sim \N\left(\muvec^{\ast}, \Sig^{\ast}\right).$$
Because the $\PMD$ is a discrete distribution, we apply a continuity correction here. For any $\xvec\in\Xset$, the corresponding reduced version is $\xvec^{\ast}$ (i.e., drop the last element of $\xvec$). We define the following $(m-1)$-dimensional hyper-cuber that is centered at $\xvec^{\ast}$,
\begin{align}\label{eqn:hyper.cube}
\mathcal{A}_{\xvec^{\ast}}=[x_1-0.5, x_1+0.5]\times\cdots\times [x_{m-1}-0.5, x_{m-1}+0.5].
\end{align}
For the NA method, the pmf is approximated as
\begin{align}\label{eqn:p.na}
p_{\NA}(\xvec)=\Pr\left(\Zvec \in \mathcal{A}_{\xvec^{\ast}} \right),
\end{align}
where $\Zvec \sim\N\left(\muvec^{\ast}, \Sig^{\ast}\right)$. In the following, we consider an error bound that gives us insights on the accuracy of the NA method.

\begin{thm}\label{thm:normal}
Let $\Xvec\sim\PMD(\Pmat)$, and we consider the reduced version $\Xvec^{\ast}$, which has mean $\mvec^{\ast}$ and non-singular variance-covariance matrix $\Sig^{\ast}$. There exists a non-singular matrix $\Cmat$ such that $\Sig^{\ast} = \Cmat\Cmat'$. For any $\xvec\in \Xset$ and its corresponding reduced version $\xvec^{\ast}$, we consider the hyper-cube $\mathcal{A}_{\xvec^{\ast}}$ as defined in \eqref{eqn:hyper.cube}. The error bound of the approximation in \eqref{eqn:p.na} is,
\begin{equation*}
|\Pr(\Xvec^{\ast} \in \mathcal{A}_{\xvec^{\ast}}) - \Pr(\Zvec \in \mathcal{A}_{\xvec^{\ast}})| \leq c(m-1)^{\frac{1}{4}} \sum_{i=1}^{n}\EE\|\Cmat^{-1}(\Ivec_{i}^{\ast}-\pvec_{i}^{\ast})\|_1^3,
\end{equation*}
where $\Zvec \sim\N\left(\muvec^{\ast}, \Sig^{\ast}\right)$, $c$ is a constant, and $\|\cdot\|$ is the $L_1$ norm.
\end{thm}

We will not provide direct proof of Theorem~\ref{thm:normal}, because it is an application of the general results in~\citeN{article}, but extended for the scenario for the PMD case. Our main purpose is to use the error bound to provide insights for the accuracy of the NA method. Intuitively, we can write,
\begin{align*}
\Xvec^{\ast} = (X_1,\dots,X_{m-1})' = \sum_{i=1}^{n} \Ivec_{i}^{\ast}= \sum_{i=1}^{n} (I_{i1},\dots,I_{i,m-1})',
\end{align*}
which can be written as the independent sum of zero mean random vector, $\Ivec_i^{\ast} - \EE(\Ivec_i^{\ast})$. Thus, the results in \citeN{article} can be applied to the PMD case.
Note that,
\begin{align*}
\sum_{i=1}^{n}\EE\|\Cmat^{-1}(\Ivec_{i}^{\ast}-\pvec_{i}^{\ast})\|_1^3=
n^{-1/2} \left\{n^{-1}\sum_{i=1}^{n}\EE\left[(\Ivec_{i}^{\ast}-\pvec_{i}^{\ast})'(n^{-1}\Sig^{\ast})^{-1}
(\Ivec_{i}^{\ast}-\pvec_{i}^{\ast})\right]^{3/2}\right\}.
\end{align*}
Thus, $\sum_{i=1}^{n}\EE\|\Cmat^{-1}(\Ivec_{i}^{\ast}-\pvec_{i}^{\ast})\|_1^3$ is of order $n^{-1/2}$. From Theorem~\ref{thm:normal}, the error bound of the NA method is of order $(m-1)^{1/4}$ and $n^{-1/2}$. We can see that, in general, the accuracy of the NA method increase in the order of $n^{-1/2}$, but it also interacts with $(m-1)$, which is the dimension of $\Xvec^{\ast}$.

%%%%%%%%%%%%%%%%%%%%%%%%%%%%%%%%%%%%%%%%%%%%%%%%%%%%%%%%%%%%%%%%%%%%%%%%%%%%%%%%%%%%%%%%%%%%%%%%%
\subsection{Simulation-Based Method}
%%%%%%%%%%%%%%%%%%%%%%%%%%%%%%%%%%%%%%%%%%%%%%%%%%%%%%%%%%%%%%%%%%%%%%%%%%%%%%%%%%%%%%%%%%%%%%%%%

Because $\Ivec_i, i=1,\dots,n$, is an $m$-dimensional vector of random indicators, it follows a multinomial distribution that has only one trial with probability $\pvec_i$. Therefore, for all $i=1,\dots, n$, one can simulate $\Ivec_i$ from Multinomial$(1, \pvec_i)$. Then $\Xvec = \sum_{i=1}^{n}\Ivec_i$ is a random sample from $\PMD$ with $\Pmat = (\pvec_1, \dots,\pvec_n)'$. We can repeat this sampling process to generate enough samples from $\Xvec \sim \PMD(\Pmat)$. Then one can use the samples to approximate the true distribution, which we denote it as $p_{\SIM}(\xvec)$. The detailed algorithm is described in \textbf{Algorithm}~\ref{algo:sim.method}. We implement the SIM method by using C++ program. The following result provides insights on the accuracy of the SIM method.

\begin{algorithm}[t]
\caption{Simulation algorithm for computing $p_{\SIM}(\xvec)$ for $\xvec \in \Xset$.}\label{algo:sim.method}
\begin{algorithmic}
\State Initialize counter $s=0$, and set $b$ for the number of repeats.
\For{$r = 1$ to $b$}
    \For{$i = 1$ to $n$}
		\State Generate $\Ivec_i$ from Multinomial$(1, \pvec_i)$
	\EndFor \State End loop for $i$
 \State Calculate $\xvec_{r}=\sum_{i=1}^{n}\Ivec_i$
	\If{$\xvec_{r}=\xvec$} \State $s=s+1$
	\EndIf
\EndFor \State End loop for $r$
\State  Obtain $p_{\SIM}(\xvec) = s/b$
\end{algorithmic}
\end{algorithm}

\begin{thm}
Consider $\Xvec\sim\PMD(\Pmat)$ and the SIM method with $b$ repeats. The total number of the elements in the support set $\Xset$ is $h=h(m,n)$. The pmf $p(\xvec)$ is estimated by $p_{\SIM}(\xvec)$ in the simulation approach. We have the following approximate expected absolute error for a single point $\xvec$,
\begin{align}\label{eqn:sim.error.single.point}
\EE|p(\xvec) - p_{\SIM}(\xvec)| \approx  \sqrt{\frac{2 p(\xvec)[1-p(\xvec)]}{\pi b}}\leq\sqrt{\frac{1}{2\pi b}}.
\end{align}
The approximate expected total absolute error,
\begin{align}\label{eqn:sim.tae}
\sum_{\xvec \in \Xset}\EE|p(\xvec) - p_{\SIM}(\xvec)| \leq \sqrt{\frac{2(h-1)}{\pi b}}.
\end{align}
\end{thm}

For a given $\xvec \in \Xset$, let $Y(\xvec)$ be the Bernoulli random variable with success probability $p(\xvec)$ (i.e., the simulated counts equal to $\xvec$). By repeating the trial for $b$ times, we obtain random variables $Y_1(\xvec),\dots,Y_b(\xvec)$. We have  $p_{\SIM}(\xvec)=\bar{Y}(\xvec)$, which is the mean of $Y_1(\xvec),\dots,Y_b(\xvec)$. By CLT,
\begin{equation*}
\sqrt{b} \left[\bar{Y}(\xvec)-p(\xvec) \right]\xrightarrow{d} \N[0,\sigma^2(\xvec)].
\end{equation*}
where $\sigma^2(\xvec) = p(\xvec)[1-p(\xvec)]$. Thus the distribution of $\bar{Y}(\xvec)$ can be approximated by a normal distribution for a large $b$. Then the expectation of absolute error for a single $p(\xvec)$ is approximated as,
\begin{equation*}
 \EE |\bar{Y}(\xvec)-p(\xvec)| = \sqrt{\frac{2}{\pi}}\sigma(\xvec) = \sqrt{\frac{2}{\pi b}p(\xvec)[1-p(\xvec)]}\leq \sqrt{\frac{1}{2\pi b}},
\end{equation*}
which is \eqref{eqn:sim.error.single.point}.

To establish \eqref{eqn:sim.tae}, let $c = \sqrt{2/(\pi b)}$. Using the inequality between arithmetic and quadratic means, we have,
\begin{align*}
& \sum_{\xvec \in \Xset}\EE |\bar{Y}(\xvec)-p(\xvec)| \approx c \sum_{r=1}^{h} \sqrt{p(\xvec)[1-p(\xvec)]}  = c h \sum_{\xvec\in\Xset}\frac{\sqrt{p(\xvec)[1-p(\xvec)]}}{h} \\
    & \leq c h \sqrt{\sum_{\xvec \in \Xset}p(\xvec)[1-p(\xvec)]/h} = c \sqrt{h} \sqrt{1-\sum_{\xvec \in \Xset} p(\xvec)^2} \\ &\leq c \sqrt{h} \sqrt{1-1/h} = c\sqrt{h-1} = \sqrt{\frac{2(h-1)}{\pi b}}.
\end{align*}

The result in \eqref{eqn:sim.error.single.point} shows that we can achieve high accuracy for a single point using the SIM method, by choosing a large $b$. However, because of the result in \eqref{eqn:sim.tae}, to control the total absolute error, the needed number of repeats grows fast as $h$ also grows fast with large $n$ and $m$.

%%%%%%%%%%%%%%%%%%%%%%%%%%%%%%%%%%%%
\section{Method Comparisons}\label{sec:Method Comparisons}
%%%%%%%%%%%%%%%%%%%%%%%%%%%%%%%%%%%%%%%%%%%%%%%%%%%%%%%%%%%%%%%%%%%%%%%%%%%%%%%%%%%%
In this section, we compare the numerical accuracy of the three methods. We also study the time efficiency of the $\dft$ method. Based on results from the accuracy and time efficiency study, we provide recommendations for the use of those methods in practice.

%%%%%%%%%%%%%%%%%%%%%%%%%%%%%%%%%%%%%%%%%%%%%%%%%%%%%%%%%%%%%%%%%%%%%%%%%%%%%%%%%%%%%%%%%%%%%%%%%
\subsection{Accuracy of the $\dft$ method}
In this section, we verify the accuracy of the $\dft$ method  through three special cases. Although there are no existing methods available to compute the pmf as the true values for large $n$ and $m$, some special cases can be considered.  The first one is binomial distribution, which is a special case of $\PMD$ when $m=2$ and all rows of $\Pmat$ are identical. The second one is the Poisson binomial distribution. The $\PMD$ becomes the Poisson binomial distribution when $m=2$, of which the pmf can be computed via the algorithm provided in \citeN{Hong2013}. The third one is $\PMD$ with small $n$ and $m$, of which the pmf can be calculated using enumeration.

The accuracy criteria used here are the maximum absolute error ($\MAE$) and the total absolute error ($\TAE$). In particular, the $\MAE$ is defined as,
$$\mathrm{\MAE} = \underset{\xvec \in \Xset}{\max}|p(\xvec) - p_{\dft}(\xvec)|.$$
which is the maximum value of the absolute differences between the true $p(\xvec)$ and the $p_{\dft}(\xvec)=p^{\ast}(\xvec^{\ast})$ computed by the $\dft$ method as shown in \eqref{eq:dft.cf.formula}. The $\TAE$ is defined as,
\begin{align*}
\mbox{\TAE} = \sum_{\xvec \in \Xset} |p(\xvec) - p_{\dft}(\xvec)|.
\end{align*}

\begin{figure}
\begin{center}
\begin{tabular}{cc}
\includegraphics[width=0.48\textwidth]{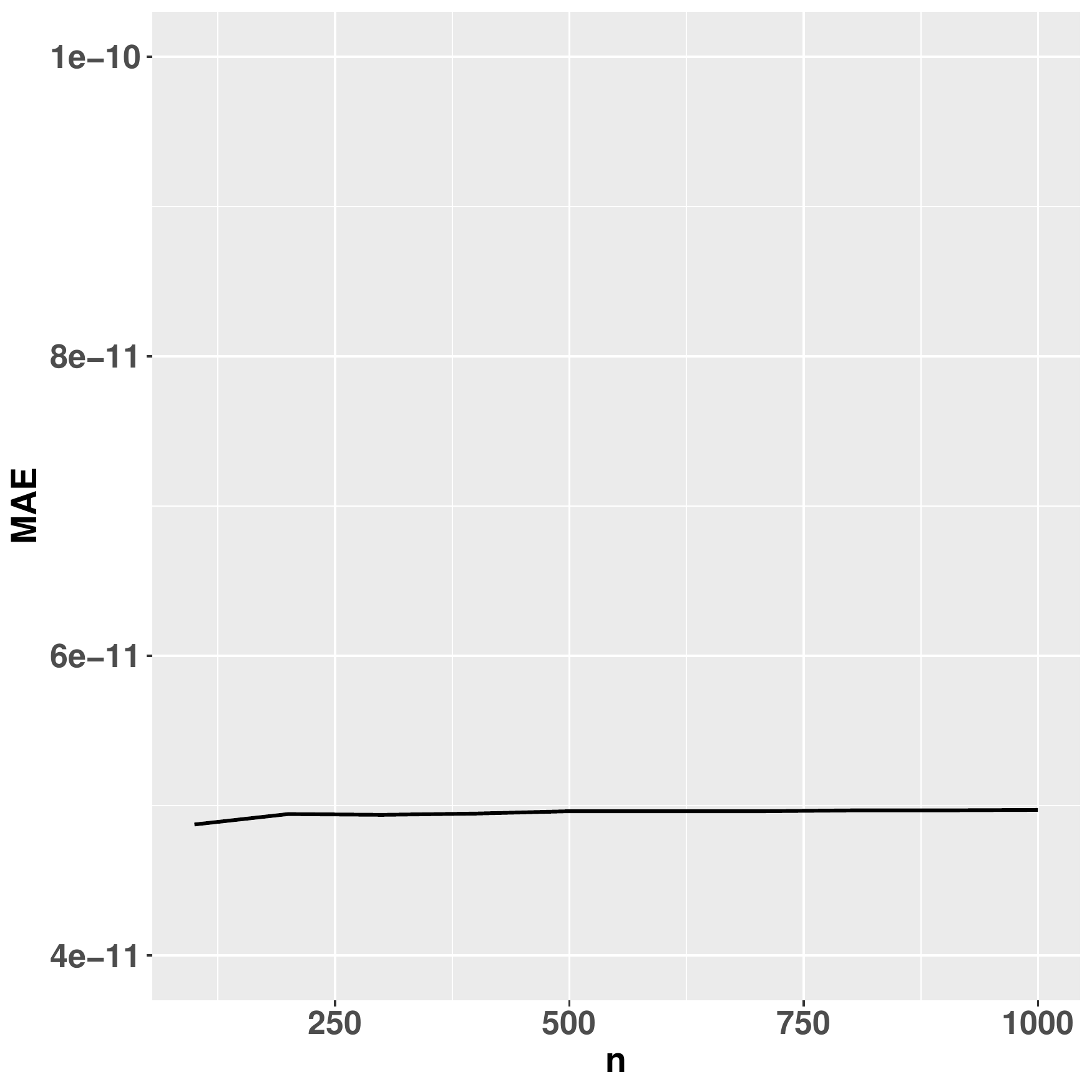} &
\includegraphics[width=0.48\textwidth]{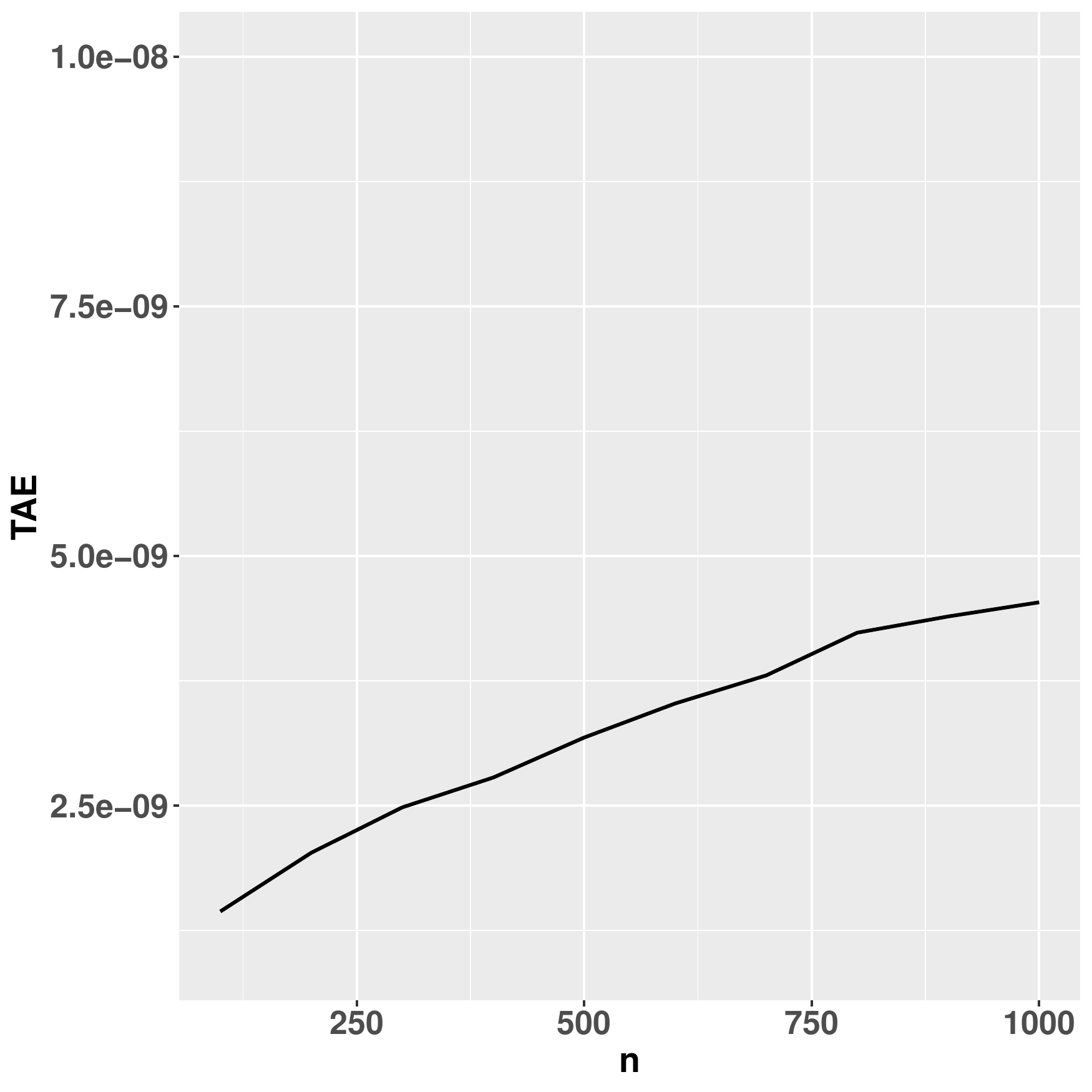} \\
(a) MAE  & (b) TAE
\end{tabular}
\end{center}
\caption{Plots of the $\MAE$ (a) and $\TAE$ (b) as a function of $n$ for the $\dft$ method using the binomial pmf as the true values.}\label{fig:mae.tae.bino}
\end{figure}

\begin{figure}
\begin{center}
\begin{tabular}{cc}
\includegraphics[width=0.48\textwidth]{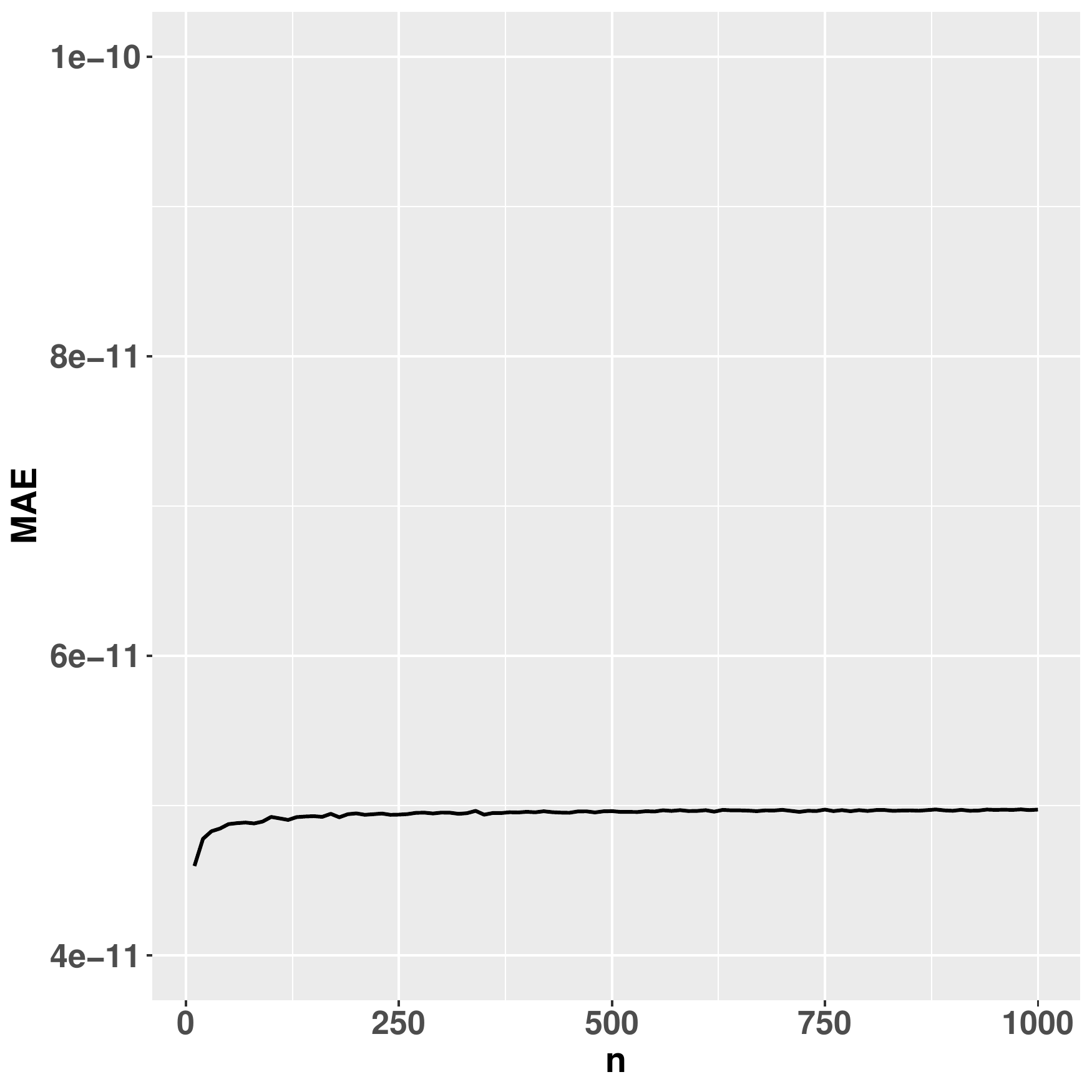} &
\includegraphics[width=0.48\textwidth]{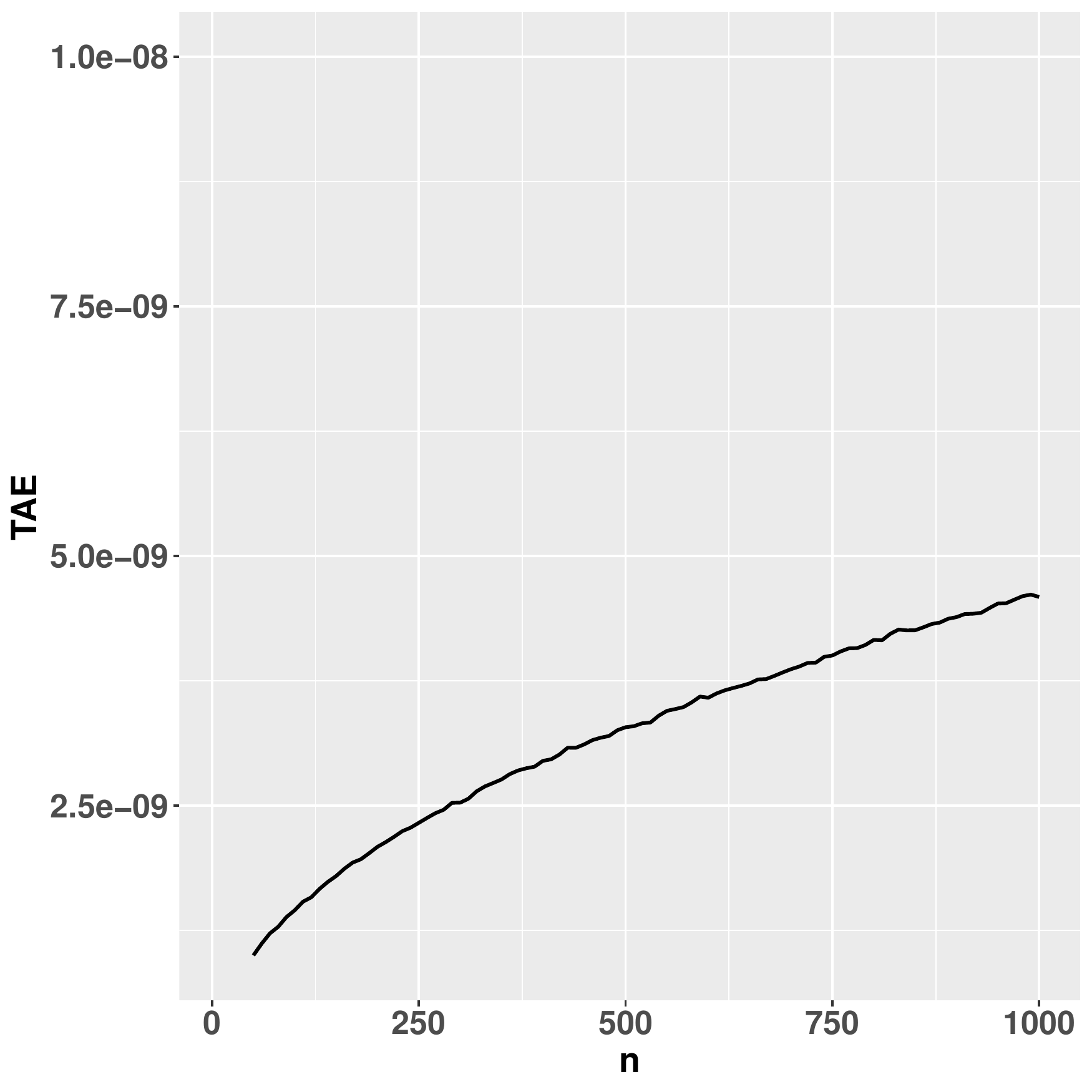} \\
(a) MAE  & (b) TAE
\end{tabular}
\end{center}
\caption{Plots of the $\MAE$ (a) and $\TAE$ (b) as a function of $n$ for the $\dft$ method using Poisson binomial pmf as the true values.}\label{fig:mae.tae.poi}
\end{figure}

In both the binomial and Poisson binomial scenarios, we randomly generate 1000 SPM, and compute the averaged $\MAE$ and $\TAE$ with $n$ ranging from 1 to 1000. The results are shown in Figures~\ref{fig:mae.tae.bino} and~\ref{fig:mae.tae.poi}. Figure~\ref{fig:mae.tae.bino} shows that the $\MAE$ is smaller than $10^{-10}$ and $\TAE$ is well controlled and is smaller than $10^{-8}$.  Figure~\ref{fig:mae.tae.poi} shows similar results when the true values are computed from the Poisson binomial distribution. The patterns of the $\MAE$ and $\TAE$ are almost the same as those in Figure~\ref{fig:mae.tae.bino}.

Further, we randomly generate ten SPM with $n$ from 2 to 5 and $m$ from 3 to 5 to test the accuracy of the $\dft$ method. We find that the method can compute the pmfs exactly the same as those computed by enumeration (e.g., as the one shown in Example~\ref{eg:simple.example}). In summary, the $\dft$ method can compute the pmf for $\PMD$ with enough accuracy for the three scenarios considered in this section.

\subsection{Accuracy of the Normal Approximation Method}
%%%%%%%%%%%%%%%%%%%%%%%%%%%%%%%%%%%%%%%%%%%%%%%%%%%%%%%%%%%%%%%%%%%%%%%%%%%%%%%%%%%%%%%%%%%%%%%%%
With the accuracy of the $\dft$ method verified, we use the pmf computed by the DFT-CF method as the true value and conduct accuracy verification for the other two methods. Accordingly, to test the accuracy of the NA method, the MAE is defined,
$$\mathrm{\MAE} = \underset{\xvec \in \Xset}{\max}|p_{\dft}(\xvec) - p_{\NA}(\xvec)|.$$

\begin{figure}%[h]
\begin{center}
\includegraphics[width=0.8\textwidth]{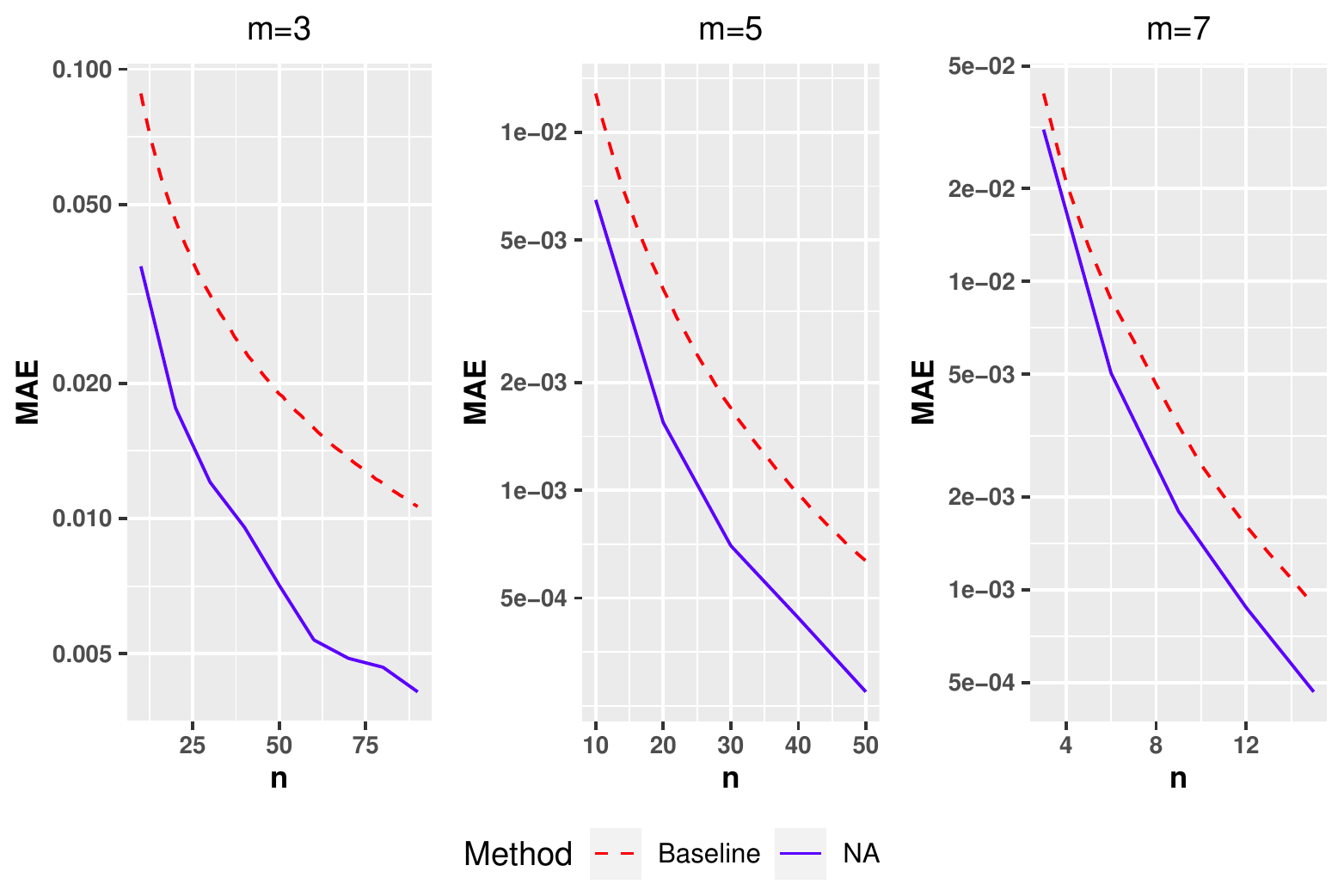}
\caption{Plot of the $\MAE$ as a function of $n$ for the NA method, compared with the baseline values, when $m=3,5,$ and 7.}\label{fig:mae.na}
\end{center}
\end{figure}

Due to computational limits, we set $m$ to be smaller than 10 and let $n$ grow to sufficiently large to test the accuracy for the NA method. For each $(n, m)$ pair, we measure the averaged $\MAE$ for 5000 randomly generated SPM and plot the curves in Figure~\ref{fig:mae.na}. As a reference, we also plot a curve that we call the baseline curve. For a given $(n,m)$, the baseline value is the largest value in the pmf for a given SPM. One can consider the baseline value as the MAE of a method that computes the pmf as zero. The baseline curve is obtained by averaging over the 5000 randomly generated SPM.

Figure~\ref{fig:mae.na} shows that, for a fixed $m$, the solid curve that represents the NA method  decreases as $n$ increases. Also as $n$ increases, the gap between the $\NA$ curve and the baseline curve becomes wider, which indicates that the accuracy of the NA method improves when $n$ becomes large. Theorem~\ref{thm:normal} suggests that the accuracy increases with the order of $n^{1/2}$, but that also interacts with $m$. With larger $m$, the accuracy increases slower as $n$ grows. This is also indicated by the results in Figure~\ref{fig:mae.na}, which shows that, for larger $m$, the gap between the NA curve and the baseline curve grows slower.

\subsection{Accuracy of the Simulation Method}
The SIM method is more convenient to use for computing the pmf at some specific points of interest because it is not time efficient to use the $\SIM$ method to compute the entire pmf. To test the accuracy of the SIM method, we consider the scenario under $m=3$ and let $n$ increase to sufficiently large. Because of time and hardware constraints, we choose to investigate the pmf at several special points in $\Xset$ to illustrate how the accuracy of the SIM method changes as the number of repeats $b$ increases.
The mode of the pmf, $\xvec_{\textrm{mode}}$, is an important point to consider. In addition, we also consider a value, denoted by $\xvec_{0.95}$, whose pmf value $p(\xvec_{0.95})$ is larger than $95\%$ of the entire pmf (i.e., $p(\xvec)$ for all $\xvec\in\Xset$). We call $\xvec_{0.95}$ the 0.95 mode of the distribution. Similarly we also consider $\xvec_{0.9}$, which is the 0.9 mode of the distribution.

Then we can use the SIM method  to compute the probability $p_{\SIM}(\xvec_{\textrm{mode}})$, $p_{\SIM}(\xvec_{0.95})$, and $p_{\SIM}(\xvec_{0.9})$. For the SIM method, we use a different criterion other than the $\MAE$ or $\TAE$. The criterion we use here is the absolute error (AE), which is defined as,
$$
\textrm{AE} = |p_{\dft}\left(\xvec_{q}\right) - p_{\SIM}\left(\xvec_{q}\right)|,
$$
where $\xvec_{q}$ can take values from $\{\xvec_{\textrm{mode}}, \xvec_{0.95}, \xvec_{0.90}\}$.

\begin{figure}%[h]
\begin{center}
\includegraphics[width=0.8\textwidth]{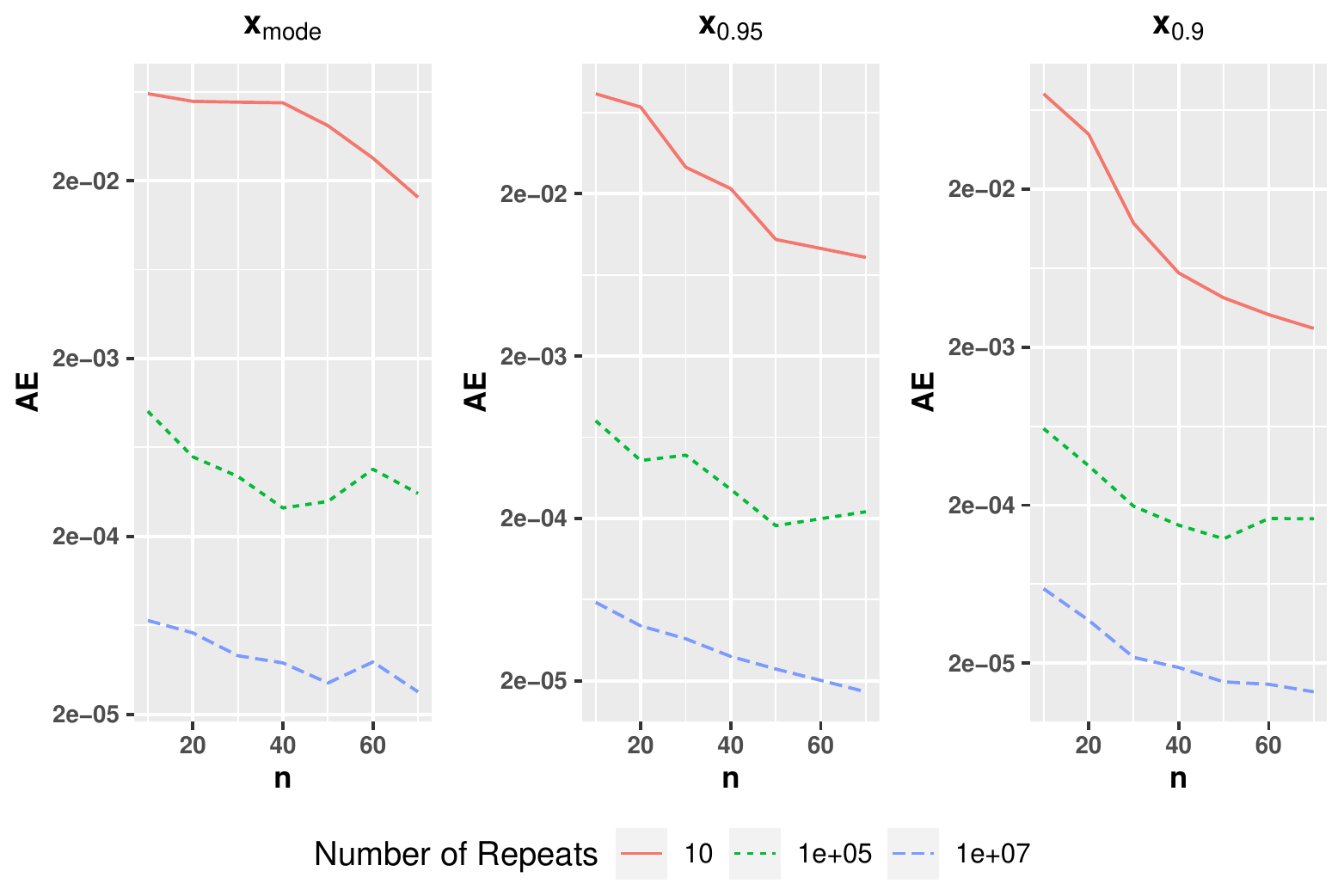}
\caption{Plot of AE as a function of $n$ for the $\SIM$ method under different number of repeats, when $m=3$.}\label{fig:accuracy.sim}
\end{center}
\end{figure}

For each $n$ from 1 to 75, we randomly generated 1000 SPM and computed the averaged AE, with the number of repeats $b$ equals to 10, $10^{5}$ and $10^{7}$. Figure~\ref{fig:accuracy.sim} plots the AE as a function of $n$ for the $\SIM$ method under different number of repeats, when $m=3$. From Figure~\ref{fig:accuracy.sim}, we can see that the AE decreases as $n$ increases. It is obvious that $b=10$ is not accurate. When $b=10^5$, the AE is between $10^{-3}$ and $10^{-4}$, while when $b=10^7$, the AE is between $10^{-4}$ and $10^{-5}$. The accuracy performance when $b$ is large is evidently better than when $b$ is small. We can tell from the plots that $b=10^7$ is a reasonable choice for the number of repeats because it provides an AE as small as $10^{-5}$.

%%%%%%%%%%%%%%%%%%%%%%%%%%%%%%%%%%%%%%%%%%%%%%%%%%%%%%%%%%%%%%%%%%%%%%%%%%%%%%%%%%%%%%%%%%%%%%%%%
\subsection{Time Efficiency of the DFT-CF method}
%%%%%%%%%%%%%%%%%%%%%%%%%%%%%%%%%%%%%%%%%%%%%%%%%%%%%%%%%%%%%%%%%%%%%%%%%%%%%%%%%%%%%%%%%%%%%%%%%
In this section, we study the computing efficiency of the DFT-CF method. Note that the DFT-CF method returns the entire pmf function (i.e., $p(\xvec)$ for all $\xvec\in\Xset$) as a result of the FFT. For example, for a $\PMD$ with $\Pmat$ of size $n \times m$, the $\dft$ method returns an $(n+1)^{m-1}$ array, although the number of non-zero points is $h(n, m)$. Note that the number $(n+1)^{m-1}$ increases fast as $m$ increases.

\begin{figure}%[h]
\begin{center}
	\includegraphics[width=0.95\textwidth]{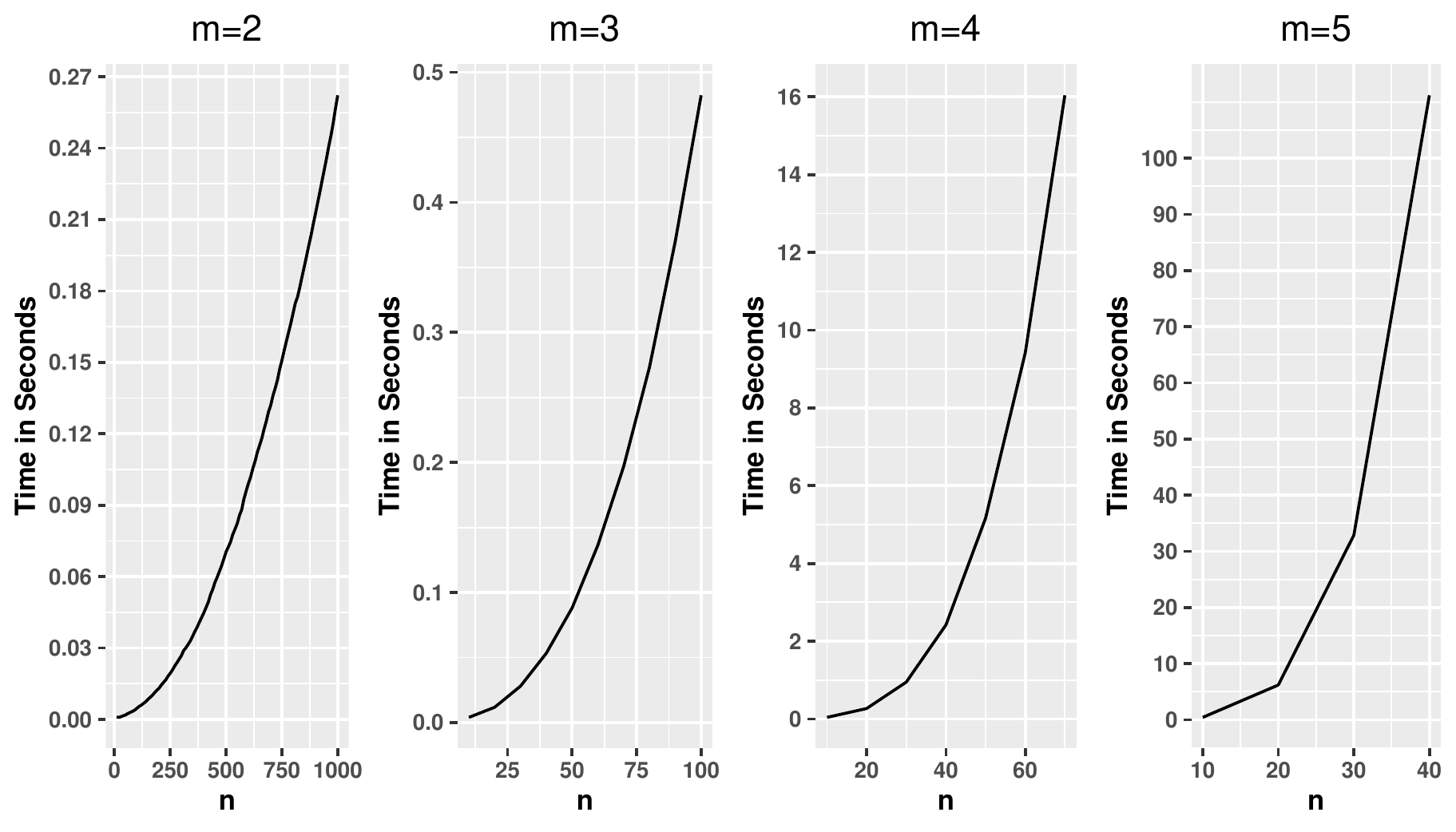}
	\caption{Plot of the computing time in seconds as a function of $n$ using the $\dft$ method, when $m=2,3,4$, and 5.}\label{fig:time.eff}
\end{center}
\end{figure}

We consider various combinations of $n$ and $m$ in the time efficiency study. For each pair of $(n, m)$, we generate 1000 SPM and the average computing time in seconds is recorded.  The system used for computing was AMD EPYC 7702 (128 cores, 2GHz) with 256GB RAM. Figure~\ref{fig:time.eff} plots the computing time in seconds as a function of $n$ using the $\dft$ method, when $m=2,3,4$, and 5.

Figure~\ref{fig:time.eff} shows that when $m$ is small (less or equal to 4), the $\dft$ method is generally fast in computing the pmf. The computing time for $n=60, m=4$ is about 16 seconds which is affordable. Even when $m=5$ and $n=40$ the time is around 100 seconds which is still acceptable. When $m$ is moderate ($8 \leq m \leq 20$) or larger, the number $(n+1)^{m-1}$ will be enormous so that the computing time is too long or the required memory may exceed the hardware limits. For example, when $m=8$ and $n=15$, the output of the $\dft$ method has $16^7$ points, and it will be both time and memory consuming.

We also want to briefly comment on the time efficiency of the NA and SIM methods. Note that both the $\NA$ and $\SIM$ methods are designed to compute some specific points of interest of the pmf, while the $\dft$ method computes the entire pmf. For the $\NA$ method, the computing time is generally small because there are efficient algorithms to compute the distribution function of the multivariate normal distribution. For the SIM method, there is always a trade-off between the accuracy and time efficiency, because the accuracy increases when the number of repeats increases, and the increase in the number of repeats results in more computing time.

%%%%%%%%%%%%%%%%%%%%%%%%%%%%%%%%%%%%%%%%%%%%%%%%%%%%%%%%%%%%%%%%%%%%%%%%%%%%%%%%%%%%%%%%%%%%%%%%%
\subsection{Practical Recommendations}
%%%%%%%%%%%%%%%%%%%%%%%%%%%%%%%%%%%%%%%%%%%%%%%%%%%%%%%%%%%%%%%%%%%%%%%%%%%%%%%%%%%%%%%%%%%%%%%%%
According to the results of the accuracy and efficiency study, we provide the following recommendations regarding the use of the three methods.

\begin{inparaitem}
\item For small $m$ (i.e., $m\leq 5$) and not large $n$, we recommend the $\dft$ method because it computes the pmf accurately and efficiently.

\item For moderate $m$ (i.e., $6\leq m \leq 20$) and small $n$, we recommend the SIM method with $b$ around $10^6$ to compute the pmf.

\item For large $n$ (i.e., cases that can not be handled by the DFT-CF method), we recommend the $\NA$ method because it can compute the pmf efficiently with enough accuracy.
\end{inparaitem}

%%%%%%%%%%%%%%%%%%%%%%%%%%%%%%%%%%%%%%%%%%%%%%%%%%%%%%%%%%%%%%%%%%%%%%%%%%%%%%%%%%%%%%
\section{Applications}\label{sec:applications}
%%%%%%%%%%%%%%%%%%%%%%%%%%%%%%%%%%%%%%%%%%%%%%%%%%%%%%%%%%%%%%%%%%%%%%%%%%%%%%%%%%%%%%

In this section, we illustrate the applications of the $\PMD$ in political science, ecological inference, and machine-learning classifications.

%%%%%%%%%%%%%%%%%%%%%%%%%%%%%%%%%%%%%%%%%%%%%%%%%%%%%%%%%%%%%%%%%%%%%%%%%
\subsection{Calculation of Voting Probability}
%%%%%%%%%%%%%%%%%%%%%%%%%%%%%%%%%%%%%%%%%%%%%%%%%%%%%%%%%%%%%%%%%%%%%%%%%
In voting scenarios, we are interested in which candidate will win the election and the probability for each candidate to win the election. The PMD can be used to answer those questions.
Suppose there are $n$ voters and $m$ candidates in an election. There will be $h(n, m)$ possible outcomes in $\Xset$. Each $\xvec$ is an $m$-dimensional vector that has $m$ elements denoting the number of votes each candidate obtains. If $\Pmat$ is provided, then we are able to compute the probability of outcomes.

For illustrations, suppose there are $n=10$ voters and $m=3$ candidates in a small scale election. Because our focus on applying the developed method to compute the PMD probability, we assume the SPM $\Pmat$ is given as follows,
\begin{equation*}
\Pmat' =
\begin{pmatrix}
 0.180 & 0.035 &0.439 &0.159 &0.350& 0.294 &0.099 & 0.102 & 0.359 &0.483\\
 0.333 & 0.348 &0.211 &0.457 &0.380& 0.696 &0.422 & 0.323 & 0.456 &0.071\\
 0.487 & 0.617 &0.350 &0.384& 0.270& 0.010 &0.479 & 0.575 & 0.185 & 0.446
\end{pmatrix}.
\end{equation*}
In practice, historical information, polls, and statistical and machine-learning methods can be used to estimate the SPM.

\begin{figure}
\begin{center}
\begin{tabular}{cc}
\includegraphics[width=0.48\textwidth]{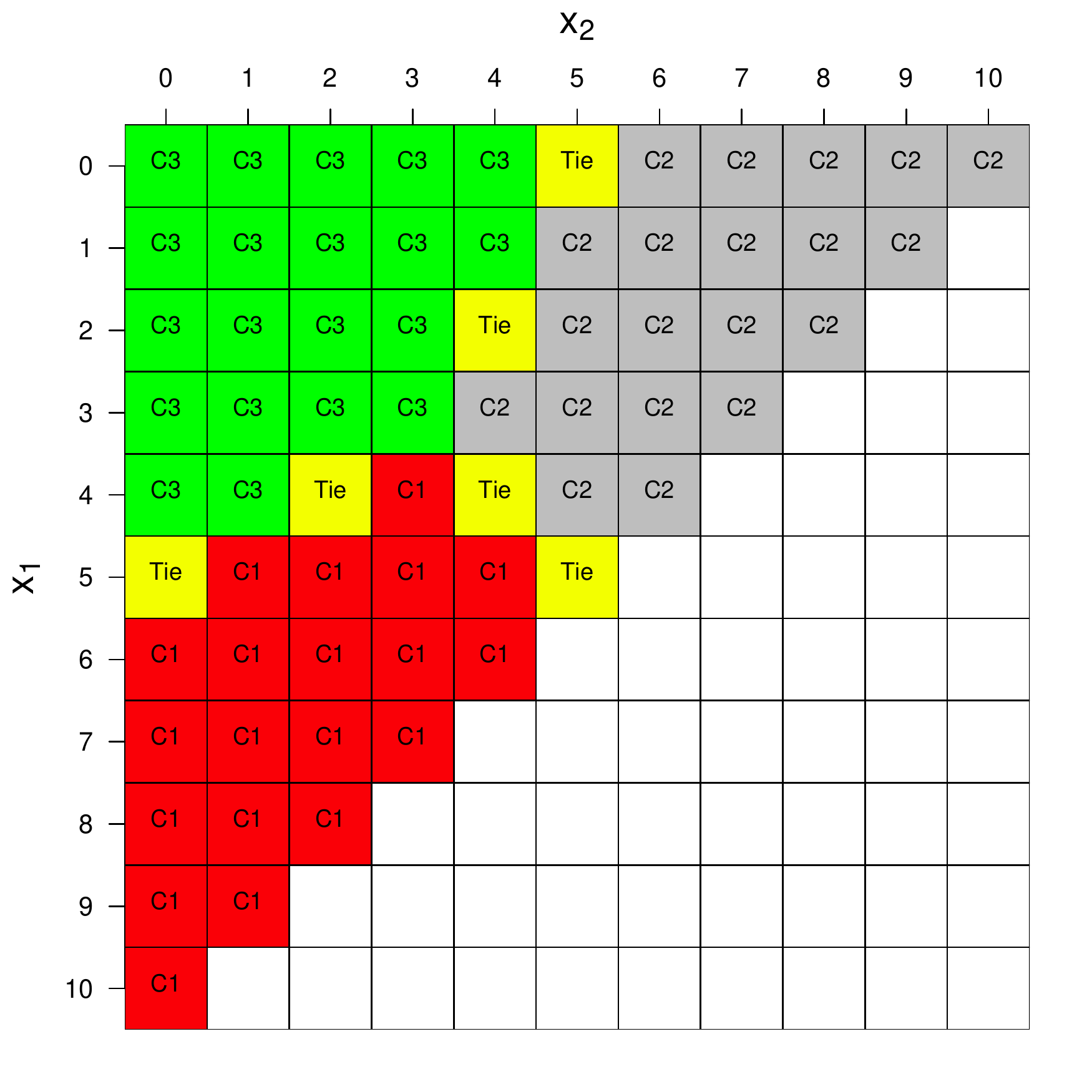}&
\includegraphics[width=0.47\textwidth]{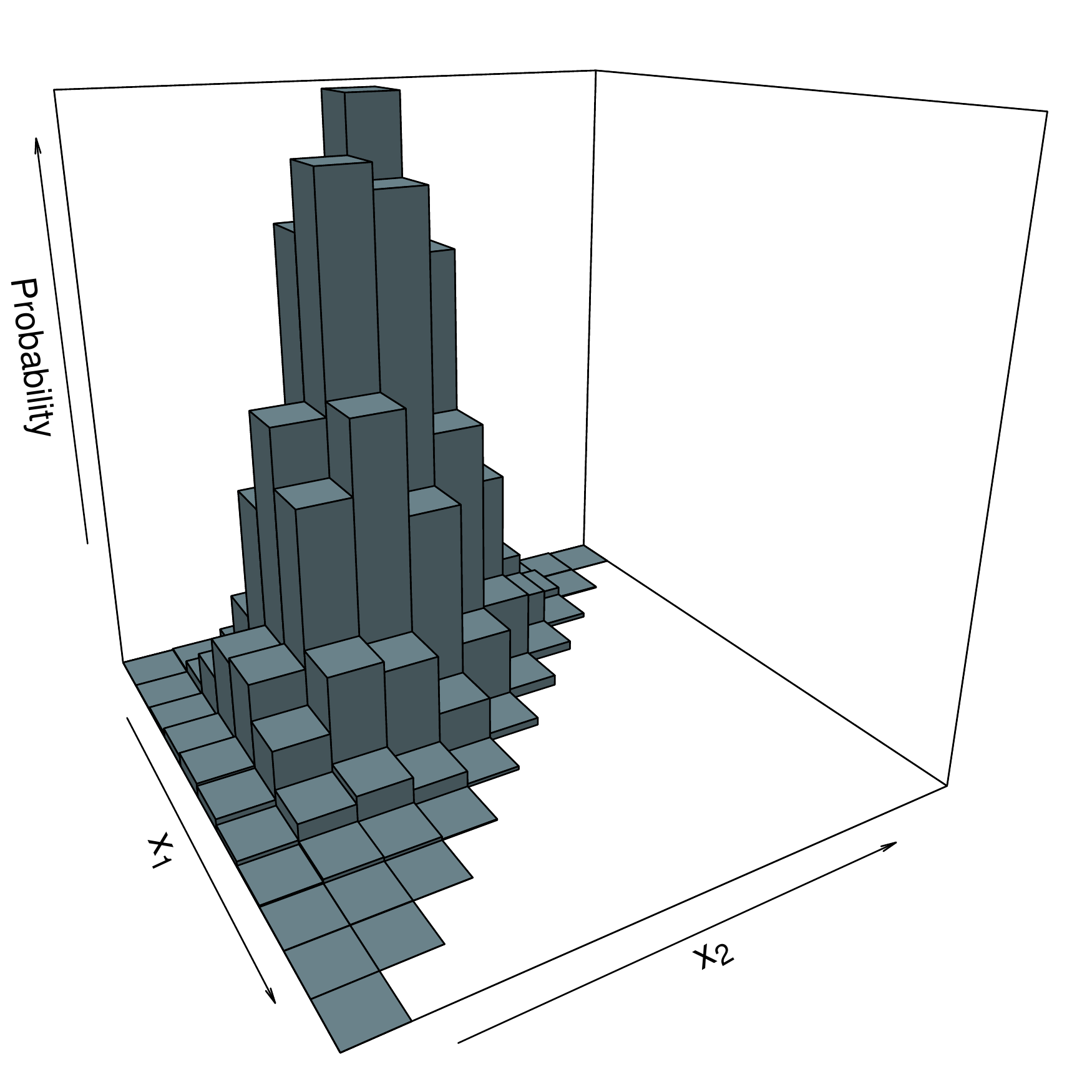}\\
(a) Winning Scenarios & (b) Plot of pmf
\end{tabular}
\caption{Plot of possible outcomes that indicate under which scenarios a candidate wins the election~(a), and the 3D barplot of pmf $p(\xvec)$~(b). The label in the square (e.g., ``C1") indicates which candidate wins the election, and ties are possible.}\label{fig:3dplot}
\end{center}
\end{figure}

Let $\xvec=(x_1, x_2, x_3)'$ be the possible counts of votes that those candidates receive. The counts follow the PMD and we use the pmf of PMD to obtain the probability that each candidate wins the election. We introduce,
$$\Xset_i = \left\{\xvec: x_i>x_j, j\neq i, \xvec \in \Xset\right\},$$
as the set of possible outcomes that candidate $i$ will win, $i=1, 2, 3$. Figure~\ref{fig:3dplot}(a) shows $\Xset_1$, $\Xset_2$ and $\Xset_3$, labeled by C1 (candidate~1), C2 (candidate~2) and C3 (candidate~3), respectively. The ``Tie" areas denote the outcomes that no one wins the election.

Figure~\ref{fig:3dplot}(b) shows the barplot of the pmf of the corresponding $\PMD$, in which the $z$ axis is the probability. The $x_1$ and $x_2$ axes show the number of votes that candidate~1 and candidate~2 receive, respectively. We can see that the mode of the pmf is $\xvec=(2, 4, 4)'$, which has the highest probability as 0.0864. Using the pmf, the probability of candidate 1 winning is,
\begin{equation*}
\Pr(\text{candidate 1 wins}) = \sum_{\xvec \in \Xset_{1}} p(\xvec) = 0.109.
\end{equation*}
Similarly, the probabilities for candidate~2 and candidate~3 winning are $0.345$ and $0.373$, respectively.

%%%%%%%%%%%%%%%%%%%%%%%%%%%%%%%%%%%%%%%%%%%%%%%%%%%%%%%%%%%%%%%%%%%%%%%%%%%%%%%%%%%%%%
\subsection{Statistical Inference for Aggregated Data}\label{sec:model.est.inf}	
Here we give an application to show how the PMD can be used to make statistical inference for aggregated data. Suppose there are $n$ individuals in an original dataset. The response is a categorical variable with $m$ levels. There are $v$ number of covariates for the response variable.

According to a certain criterion, those $n$ individuals are grouped into $h$ groups, and the corresponding group size is $n_i,i=1,\dots,h$. Due to privacy protection, proprietary sensitive information, or other practical reasons, only the aggregated counts for the response variable of the group are reported. That is, we only have the counts for the aggregated response variable of the $i$th group, which is denoted by
$\xvec_i = (x_{i1}, \dots,x_{im})'$, $i=1, \dots, h$, and $x_{ij}$ is the counts of the individuals with the $j$th level of the response variable within group $i$.  The individual level covariate information is still available for analysis. Let $\Gmat_i$ be the matrix that contains all covariate information for the $i$th group. Then $\Gmat_i = (\gvec_{i 1}, \dots, \gvec_{ij},\dots,\gvec_{i n_i})'$ is an $n_i \times (v+1)$ matrix and $\gvec_{ij}$ contains the covariate information for individual $j$ in group $i$, with the first element of $\gvec_{ij}$ being 1 for the intercept term. The statistical inference can only be based on the aggregated data.

To model the relationship between the aggregated counts and the covariates, logistic-type models can be used. Let $\Pmat_i = (p_{ijk})$ be the SPM for the $i$th group, $i = 1, \dots, h$, $j = 1,\dots,n_i$, and $k = 1,\dots, m$. Using the softmax function, the probability $p_{ijk}$ is linked to the covariates through,
\begin{align}\label{eqn:pijk.logistic.model}
    p_{i j k} = \frac{\exp{\left(\gvec_{ij}' \betavec_{k}\right)}}{1 + \sum_{k=1}^{m-1}\exp{\left( \gvec_{ij}' \betavec_{k} \right)}},
    \quad k \neq m, \quad \text{and } \quad
    p_{i j m} = \frac{1}{1 + \sum_{k=1}^{m-1}\exp{\left( \gvec_{ij}' \betavec_{k} \right)}},
\end{align}
for all $i=1,\dots, h, j=1,\dots, n_i, k=1,\dots, m$. Here, we set category $m$ as the baseline, and $\betavec_j = \left(\beta_{j0},\dots,\beta_{jv}\right)', j=1,\dots,m-1$. Let $\betavec = \left(\betavec_1, \dots, \betavec_{m-1}\right)'$ be an $(m-1) \times (v+1)$ matrix for the regression coefficients.

Let $\Xvec_i$ be the random vector for the aggregated counts for group $i$ and $\Xvec_i\sim\PMD(\Pmat_i)$. The probability of observing $\Xvec_i = \xvec_i$ is $p(\xvec_i)$, which can be computed by the proposed methods. Thus, the log-likelihood for $\betavec$ based on aggregated data from all groups is,
\begin{align}\label{eqn:log.lik.agg.dat}
\loglik(\betavec) = \sum_{i=1}^{h}\loglik_i(\betavec) = \sum_{i=1}^{h}\log[p(\xvec_i)].
\end{align}
The estimate of $\betavec$, denoted by $\wh{\betavec}$, can be obtained by maximizing the log-likelihood function in \eqref{eqn:log.lik.agg.dat}. Substituting $\wh{\betavec}$ into \eqref{eqn:pijk.logistic.model}, we obtain an estimate of the SPM $\Pmat_i$, denoted by $\wh{\Pmat}_i$, $i=1, \dots, h$.

We apply model~\eqref{eqn:pijk.logistic.model} to the
ai4i 2020 Predictive Maintenance Dataset ``ai4i'' (\citeNP{Dua:2019}). The dataset ``ai4i'' is a machine failure dataset that reflects predictive maintenance data encountered in the industry. The data consist of 10000 products (rows) and covariates, including product type, tool wear, air temperature, and rotational speed.  To obtain aggregated data, we divide the dataset into 708 groups based on the combinations of product type and tool wear.

For each product, we consider three failure statuses, which is the response. The first failure status is related to tool wear and strain, which is denoted by category~1. The second failure status is related to power and heating, which is denoted by category~2. The third failure status includes non-failure product units and random failures, which is denoted by category~3. The total counts for the three categories are 123, 189, and 9688, respectively. After data aggregation, we obtain the counts for the three categories for each of the 708 groups. For an illustration of including covariates in the model, we consider standardized air temperature and rotational speed as the covariates in \eqref{eqn:pijk.logistic.model}.

Table~\ref{tab:beta} lists the maximum likelihood estimates for $\betavec$, and the corresponding standard errors and 95\% confidence intervals, based on ``ai4i'' aggregated data. With the estimates, we can estimate $\Pmat_i$ for group $i$, denoted by $\wh{\Pmat}_i$, $i=1,\dots,h$. We demonstrate the computing of $\wh{\Pmat}_i$ by using the 1st group and 5th group as examples. The covariate matrix $\Gmat_1$ and $\Gmat_5$ for the 1st and 5th groups are,
\begin{align*}
{\Gmat}_1 = \begin{pmatrix}
1 &   -0.952     &              \hphantom{-}0.068\\
1 &          -1.202  &         -0.004\\
1 &          -1.202    &       -0.540 \\
\vdots & \vdots & \vdots \\
1 &  -0.602 & \hphantom{-}0.213\\
\end{pmatrix}_{34\times 3}, \quad
    {\Gmat}_5= \begin{pmatrix}
1   &      -0.902     &       -0.729\\
1 &         -0.552  &          -0.495\\
1 &         -1.002&            -0.138\\
\vdots & \vdots & \vdots \\
1 &        -0.702          &  -1.181\\
    \end{pmatrix}_{32\times 3},
\end{align*}
respectively. The corresponding aggregated response variables are $\xvec_1=\left(0, 1, 33\right)'$ and $\xvec_5=\left(0, 1, 31 \right)'$. The first two columns of $\wh{\Pmat}_1$ and $\wh{\Pmat}_5$ can be computed via $\Gmat_1 \wh{\betavec}$ and $\Gmat_5 \wh{\betavec}$. Thus, the $\wh{\Pmat}_1$ and $\wh{\Pmat}_5$ can be obtained as,
\begin{align*}
\wh{\Pmat}_1 = \begin{pmatrix}
0.0085 & 0.0004 & 0.9911\\
0.0078 & 0.0004 & 0.9918\\
0.0079 & 0.0021 & 0.9900\\
\vdots & \vdots & \vdots \\
 0.0096 & 0.0003 & 0.9901\\
\end{pmatrix}_{34\times 3}, \quad
    \wh{\Pmat}_5= \begin{pmatrix}
0.0089 & 0.0055 & 0.9856\\
0.0101&  0.0039 & 0.9860\\
0.0084 & 0.0008 & 0.9908\\
\vdots & \vdots & \vdots \\
0.0096 & 0.0283 & 0.9621\\
    \end{pmatrix}_{32\times 3}.
\end{align*}
Then, the estimated probability of observing the counts are $p(\xvec_1) = 0.222$ and $p(\xvec_5) = 0.232$.

\begin{table}
\begin{center}
\caption{Maximum likelihood estimates for $\betavec$, and the corresponding standard errors (SE) and 95\% confidence intervals (CIs), based on the ``ai4i'' aggregated data.}\label{tab:beta}
\begin{tabular}{crrrrccrrrr}\hline\hline
\multicolumn{5}{c}{Category 1}&&\multicolumn{5}{c}{Category 2}\\\cline{1-5}\cline{7-11}
\multirow{2}{*}{Para.}&\multirow{2}{*}{est.} &\multirow{2}{*}{SE}&\multicolumn{2}{c}{ 95\% CI}&&
\multirow{2}{*}{Para.}&\multirow{2}{*}{est.} &\multirow{2}{*}{SE}&\multicolumn{2}{c}{ 95\% CI}\\\cline{4-5}\cline{10-11}
&&& Lower & Upper&&&&& Lower & Upper\\\hline
$\beta_{10}$& $-$4.401 & 0.019 & $-$4.439 & $-$4.363 && $\beta_{20}$ & $-$6.484  & 0.045  & $-$6.572  & $-$6.396 \\
$\beta_{11}$& 0.374 & 0.023 &  0.329 & 0.419 && $\beta_{21}$& 1.125 &0.057& 1.013 & 1.236 \\
$\beta_{12}$& $-$0.044 & 0.023 & $-$0.090 & 0.002 && $\beta_{22}$&  $-$3.172 & 0.040 & $-$3.251& $-$3.093\\\hline\hline
\end{tabular}
\end{center}

\end{table}

%%%%%%%%%%%%%%%%%%%%%%%%%%%%%%%%%%%%%%%%%%%%%%%%%%%%%%%%%%%%%%%%%%%%%%%%%%%%%%%%%%%%%%
\subsection{Uncertainty Quantification in Classification}

In this section, we illustrate the use of the PMD in classification problems under machine-learning settings. In a classification problem with multiple labels, the probability that a unit belongs to each class is computed for each unit in the test set. Using a soft classifier, the predicted class is randomly assigned according to the predicted probabilities, leading to randomness in the confusion matrix. Suppose we have $m$ classes, and the $i$th observation in the test set has the predicted probability vector for each class as $\pvec_i = (p_{i1},\dots, p_{im})'$, $i=1,\cdots, n$, where $n$ is the number of observations in the test set. Then the predicted class for the observation will be a random draw based on the probability of success $\pvec_i$.

Without loss of generality, one can sort the rows in the test set by the true class level. Then the output probability matrix from the classifier can be denoted as $(\Pmat_1', \cdots, \Pmat_j', \cdots, \Pmat_m')'$, which is an $n\times m$ matrix with the $i$th row as $\pvec_i$. Here, $\Pmat_j, j=1, \cdots, m$ is the output probability matrix for the observations with true class $j$ and is of size $n_j\times m$ with $n_j$ denotes the number of observations with true class $j$.

The confusion matrix is often used to quantify the accuracy of the classifier. Table~\ref{tab:cm} shows the layout of the confusion matrix, which is an $m\times m$ matrix. For each observation, it may fall within one of those $m^2$ cells in the confusion matrix. The vector for the counts $\Xvec=(\Xvec_1', \cdots,\Xvec_m')'$ follows a PMD with $m^2$ categories, where $\Xvec_k=(X_{1k}, \cdots, X_{mk})'$ and $X_{jk}$ is the counts of observations that fall within the $(j, k)$th cell. Thus, the randomness of the soft classifier makes the PMD a suitable distribution to characterize the distribution of the counts in the confusion matrix. Then we can use the PMD to quantify the uncertainty in the confusion matrix.

\begin{table}
\caption{Illustration of confusion matrix for classification with $m$ classes. The $X_{jk}$ gives the counts out of the $n$ observations in the test set, which falls within the $(j, k)$th cell.}
\label{tab:cm}
\centering
\begin{tabular}{cc|cccc}\hline\hline
                              & & \multicolumn{4}{c}{True Class}\\\cline{3-6}
                              & & 1 & 2 & \dots & $m$\\ \hline
\multicolumn{1}{c|}{}         & 1 & $X_{11}$ & $X_{12}$ & \dots & $X_{1m}$ \\
\multicolumn{1}{c|}{Predicted}& 2 & $X_{21}$ & $X_{22}$ & \dots & $X_{2m}$\\
\multicolumn{1}{c|}{Class} &\vdots& \vdots & \vdots & $\ddots$ & \vdots\\
\multicolumn{1}{c|}{}         &$m$ & $X_{m1}$& $X_{1m}$ & \dots & $X_{mm}$\\ \hline\hline
\end{tabular}
\end{table}

For an observation with true class $k$, it can only contribute a count to those cells in the $k$th column of the confusion matrix as shown in Table~\ref{tab:cm}. Thus, the SPM for $\Xvec$ is a block diagonal matrix as follows,
$$
\Pmat=
\begin{pmatrix}
  \Pmat_1  & \zeromat & \cdots & \zeromat \\
  \zeromat & \Pmat_2  & \cdots & \zeromat \\
  \vdots   & \vdots   & \ddots & \vdots   \\
  \zeromat & \zeromat & \cdots & \Pmat_m  \\
\end{pmatrix}.
$$
Then the counts in the confusion matrix follows the $\PMD$ with $\Pmat$. That is, $\Xvec\sim\PMD(\Pmat)$.  By using Proposition~\ref{prop:Pmat.product}, the computing of the pmf of $\Xvec$ can be simplified as,
\begin{align}\label{eqn:confusion.matrix.pmf}
p(\xvec)= \Pr(\Xvec=\xvec)=\Pr(\Xvec_{1}=\xvec_{1}, \dots, \Xvec_{m}=\xvec_{m})=  \prod_{j=1}^{m}\Pr(\Xvec_{j} = \xvec_{j}).
\end{align}
Here, $\xvec=(\xvec_1', \cdots,\xvec_m')'$ is partitioned in the same way as in $\Xvec=(\Xvec_1', \cdots,\Xvec_m')'$. Uncertainty quantification can be done based on the pmf.

As an illustration, we consider an Electroluminescence (EL) image classification example in a photovoltaic (PV) reliability study. The EL image is an important data type that reveals information about the PV health status. Because disconnected parts in the PV module do not irradiate, the darker areas in EL images indicate defective cells. The EL imaging is a non-destructive technology that can provide a visual inspection of solar panels. More details on EL image for PV inspections can be found at \shortciteN{Buerhop2018}, \shortciteN{Deitsch2019}, and \shortciteN{Deitsch2021}. The dataset is available online at \shortciteN{elpv2018}. In total there are 2624 images. All images are preprocessed with respect to size and are eliminated distortion induced by the camera lens used to capture the EL images. Each image is manually labeled with its probability of defectiveness, which is one of four values, 0, 1/3, 2/3, and 1 and we label them as categories 1, 2, 3, and 4, respectively. That is $m=4$ in this application.

\begin{figure}
\begin{center}
	\includegraphics[width=0.75\textwidth]{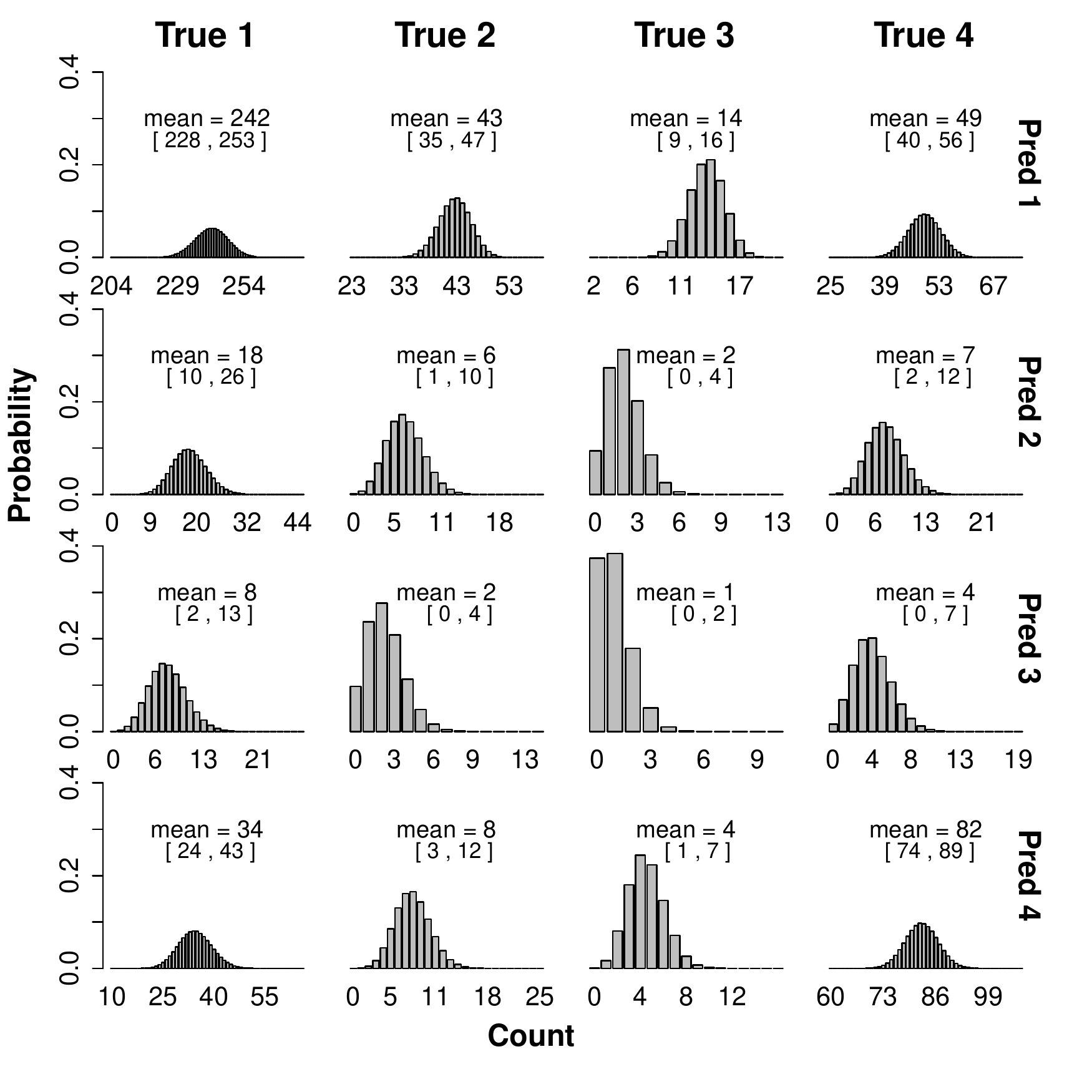}
	\caption{The barplots show the marginal pmfs for the counts in the cells of the confusion matrix. We also present the mean and 95\% naive prediction interval for the marginal counts.}
	\label{fig:confusion.hist}
\end{center}
\end{figure}

We split our data into training data (80\%) and test data (20\%), then train a convolution neural network (CNN) model on the training set (e.g., \citeNP{Goodfellow-et-al-2016}). The input predictors of the CNN model are 8-bit grayscale image data with $300\times 300$ pixels, and the output is the label of each image. The CNN model includes a convolutional layer, a fully-connected layer and an output layer. At the convolutional layer, convolutions of the input image data with kernel size $3 \times 3$ are implemented. These kernels extract features from the image data and the resulting features are fed into the fully-connected layer with the Relu activation function. Then at the fully-connected layer, the softmax activation function is applied to obtain the probability that the observation belongs to each class. As an example, Table~\ref{tbl:elpmat} provides a subset of the CNN model output.

\begin{table}%[ht]
\begin{center}
\caption{An example of the probability vectors of a subset in the testing set from the trained CNN model.}\label{tbl:elpmat}
\begin{tabular}{c|cccc}
  \hline\hline
\backslashbox{\small ID}{\small Class} & 1 & 2 & 3 & 4 \\
  \hline
  1 & 0.923 & 0.037 & 0.011 & 0.030 \\
  2 & 0.074 & 0.080 & 0.051 & 0.795 \\
  3 & 0.000 & 0.002 & 0.001 & 0.998 \\
  4 & 0.917 & 0.054 & 0.006 & 0.023 \\
  5 & 0.958 & 0.024 & 0.007 & 0.011 \\
  6 & 0.899 & 0.035 & 0.013 & 0.053 \\
   \hline\hline
\end{tabular}
\end{center}
\end{table}

In this application, we are interested in the uncertainty of the confusion matrix. We can obtain the joint pmf by using \eqref{eqn:confusion.matrix.pmf} and then the marginal pmf for the counts in each cell, which is $\Pr(X_{jk}=x_{jk})$. The barplots in Figure~\ref{fig:confusion.hist} show the estimated marginal pmfs for $X_{jk}$'s. For example, the barplot in the (1,1) panel represents the scenario that the true class is 1 and the predicted class is also 1. The $x$-axis represents the possible counts belonging to this cell, which ranges from 0 to the number of true class 1 units (i.e., $n_1$). The $y$-axis shows the corresponding probability. With the estimated pmf, we can compute the mean counts that fall in this cell. Using the estimated marginal cumulative distribution function (cdf), we can also obtain a $95\%$ naive prediction interval (i.e.,  ignore the uncertainty in the estimator of $\pvec_i$) by taking the 0.025 and 0.975 quantiles of the cdf. In this way, we can provide a way to quantify the uncertainty in the confusion matrix.

%%%%%%%%%%%%%%%%%%%%%%%%%%%%%%%%%%%%
\section{Illustrations of the R Package}\label{sec:rpackage}
%%%%%%%%%%%%%%%%%%%%%%%%%%%%%%%%%%%%%%%%%%%%%%%%%%%%%%%%%%%%%%%%%%%%%%%%%%%%%%%%%%%%%%%%%%%%%%%%%
We develop an R package ``PoissonMultinomial'' (\shortciteNP{R-PMD}) that computes the probability functions for the $\PMD$ using the methods described by this paper. The package provides functions to compute the pmf, cdf, and to generate random numbers from the PMD. There are three major functions in the package. In particular, the $\code{dpmd}$ is a function for computing the pmf, the $\code{ppmd}$ is for computing the cdf, and the $\code{rpmd}$ is for generating random numbers.

The user has to specify the $\Pmat$ so that the $\PMD$ can be determined. Also, users can specify a method to compute the pmf or cdf. If the method is not specified, the $\dft$ method is used by default. With unspecified $\xvec$, $\code{dpmd}$ automatically computes the entire pmf using the $\dft$ method and the output be a multi-dimensional array. If the user inputs $\xvec$, the output of $\code{dpmd}$ is the pmf at $\xvec$. Notice that only the $\dft$ method can automatically compute the entire pmf and it is the most efficient way for doing that. The function $\code{ppmd}$ uses the same method as $\code{dpmd}$ to compute the cdf.

The following give examples of using $\code{dpmd}$. First, $\code{pp}$ is an input matrix that specifies a $\PMD$. For example,
\begin{verbatim}
  > pp=matrix(c(0.1, 0.1, 0.1, 0.7,
              0.1, 0.3, 0.3, 0.3,
              0.5, 0.2, 0.1, 0.2),
              byrow=T, ncol=4, nrow=3)
\end{verbatim}
and $\xvec=(0,0,1,2)'$. Note that the corresponding $\xvec^{\ast}$ is $(0,0,1)'$. Then the code of using $\code{dpmd}$ to compute pmf is given as follows.
\begin{verbatim}
  > dpmd(pmat = pp)
  > dpmd(pmat = pp, xmat = x)
  > dpmd(pmat = pp, xmat = x, method = "NA" )
  > dpmd(pmat = pp, method = "SIM", B = 1e3)
  > dpmd(pmat = pp, xmat = x, method = "SIM", B = 1e3)
\end{verbatim}
The first line computes the entire pmf, the second one computes the pmf for the given $\xvec$, and the rest of the lines do a similar task using the $\NA$ and $\SIM$ methods.

For this example, the output of $\code{dpmd}$ is a $4 \times 4 \times 4$-dimensional array because the $\Pmat$ is of size $3 \times 4$, which is listed as follows and is named as $\code{res}$.
\begin{verbatim}
 > res <- dpmd(pmat=pp)
 > res
  , , 1                              , , 2
        [,1]  [,2]  [,3]  [,4]             [,1]  [,2]  [,3] [,4]
  [1,] 0.042 0.090 0.054 0.006       [1,] 0.069 0.084 0.015    0
  [2,] 0.125 0.148 0.023 0.000       [2,] 0.138 0.042 0.000    0
  [3,] 0.052 0.022 0.000 0.000       [3,] 0.021 0.000 0.000    0
  [4,] 0.005 0.000 0.000 0.000       [4,] 0.000 0.000 0.000    0
  , , 3                              , , 4
        [,1]  [,2] [,3] [,4]               [,1] [,2] [,3] [,4]
  [1,] 0.030 0.012    0    0         [1,] 0.003    0    0    0
  [2,] 0.019 0.000    0    0         [2,] 0.000    0    0    0
  [3,] 0.000 0.000    0    0         [3,] 0.000    0    0    0
  [4,] 0.000 0.000    0    0         [4,] 0.000    0    0    0
\end{verbatim}
When extracting results from the R array $\code{res}$, one needs to keep in mind that the index of an R array starts with 1, while each element of $\xvec^{\ast}$ starts with 0. For example, $\code{res[3,2,1]}=0.22$ corresponds to $p^{\ast}(\xvec^{\ast})=0.22$ at $\xvec^{\ast}=(2,1,0)'$. For another example, if one wants to find $p^{\ast}(\xvec^{\ast})$ at $\xvec^{\ast}=(0,3,0)'$, the corresponding R result is $\code{res[1,4,1]}=0.006$.

%%%%%%%%%%%%%%%%%%%%%%%%%%%%%%%%%%%%%%%%%%%%%%%%%%%%%%%%%%%%%%%%%%%%%%%%%%%%%%%%%%%%%%%%%%%%%%%%%
\section{Conclusions and Areas for Future Research}\label{sec:conclusion}
%%%%%%%%%%%%%%%%%%%%%%%%%%%%%%%%%%%%%%%%%%%%%%%%%%%%%%%%%%%%%%%%%%%%%%%%%%%%%%%%%%%%%%%%%%%%%%%%%

In this paper, we describe the PMD and explore some useful properties of PMD. We develop three methods that can be useful for computing the pmf of PMD. The $\dft$ method is an exact method, the SIM method is a simulation method, and the $\NA$ method is an approximate method. The accuracy and efficiency of those methods are studied under various scenarios. We recommend using the $\dft$ method when $m$ is small, using the $\SIM$ method when $m$ is moderate and $n$ is small, and using the $\NA$ method when $n$ is large. We also implement the three methods in an R package.

We also apply the developed methods to various areas. In political science, we show an example in an election scenario that uses $\PMD$ to compute the probabilities of possible election results. In ecological inference, we build a logistic-type model and use $\PMD$ to compute the likelihood for aggregated data. In classification, we train a CNN model and use $\PMD$ to quantify the uncertainty in the confusion matrix.

However, there are still some topics that remain to be explored. The computing speed of the $\dft$ method could be improved using more efficient Fourier transform algorithms (e.g., the convolution scheme used in \citeNP{BiscarriZhaoBrunner2018}) or one could find a way to compute only the $h(n, m)$ possible outcomes rather than $(n+1)^{m-1}$ probability mass points, which contains a large number of points with values equal to 0. The $\SIM$ method is time-consuming, although it can compute some cases that the $\dft$ method is unable to. Until now, we are still unable to compute the pmf of $\PMD$ method when $m$ is large due to the computational limit of machines, and this area remains to be challenging. In our current methods in quantifying the uncertainty in the confusion matrix, we ignore the uncertainty in the estimator of $\pvec_i$. In the future, it will also be interesting to incorporate the uncertainty in the estimator of $\pvec_i$ into the prediction interval.

%%%%%%%%%%%%%%%%%%%%%%%%%%%%%%%%%%%%%%%%%%%%%%%%%%%%%%%%%%%%%%%%%%%%%%%%%%%%%%%%%%%%%%%%%%%%%%%%%%%%%%%%%%%%%%%%%%%%
\section*{Acknowledgments}
The authors acknowledge the Advanced Research Computing program at Virginia Tech for providing computational resources.

%%%%%%%%%%%%%%%%%%%%%%%%%%%%%%%%%%%%%%%%%%%%%%%%%%%%%%%%%%%%%%%%%%%%%%%%%%%%%%%%%%%%%%%%%%%%%%%%%%%%%%%%%%%%%%%%%


\begin{thebibliography}{}

\bibitem[\protect\citeauthoryear{Akter, Moon, and Kwon}{Akter
  et~al.}{2019}]{akter2019double}
Akter, L.~A., I.~Moon, and G.-R. Kwon (2019).
\newblock Double random phase encoding with a {Poisson}-multinomial
  distribution for efficient colorful image authentication.
\newblock {\em Multimedia Tools and Applications\/}~{\em 78}, 14613--14632.

\bibitem[\protect\citeauthoryear{Bentkus}{Bentkus}{2005}]{article}
Bentkus, V. (2005).
\newblock A {Lyapunov}-type bound in {$R^d$}.
\newblock {\em Theory of Probability and Its Applications\/}~{\em 49},
  311--323.

\bibitem[\protect\citeauthoryear{Biscarri, Zhao, and Brunner}{Biscarri
  et~al.}{2018}]{BiscarriZhaoBrunner2018}
Biscarri, W., S.~D. Zhao, and R.~J. Brunner (2018).
\newblock {A simple and fast method for computing the {Poisson} binomial
  distribution function}.
\newblock {\em Computational Statistics \& Data Analysis\/}~{\em 122}, 92--100.

\bibitem[\protect\citeauthoryear{Buerhop-Lutz, Deitsch, Maier, Gallwitz,
  Berger, Doll, Hauch, Camus, and Brabec}{Buerhop-Lutz
  et~al.}{2018}]{Buerhop2018}
Buerhop-Lutz, C., S.~Deitsch, A.~Maier, F.~Gallwitz, S.~Berger, B.~Doll,
  J.~Hauch, C.~Camus, and C.~J. Brabec (2018).
\newblock A benchmark for visual identification of defective solar cells in
  electroluminescence imagery.
\newblock In {\em European PV Solar Energy Conference and Exhibition (EU
  PVSEC)}, pp.\  1287 -- 1289.

\bibitem[\protect\citeauthoryear{Cheng, Diakonikolas, and Stewart}{Cheng
  et~al.}{2017}]{Cheng2017PlayingAG}
Cheng, Y., I.~Diakonikolas, and A.~Stewart (2017).
\newblock Playing anonymous games using simple strategies.
\newblock In {\em Proceedings of the Twenty-Eighth Annual ACM-SIAM Symposium on
  Discrete Algorithms (SODA), DOI: 10.1137/1.9781611974782.40}.

\bibitem[\protect\citeauthoryear{Daskalakis, Kamath, and Tzamos}{Daskalakis
  et~al.}{2015}]{Daskalakis2015OnTS}
Daskalakis, C., G.~Kamath, and C.~Tzamos (2015).
\newblock On the structure, covering, and learning of {Poisson} multinomial
  distributions.
\newblock {\em 2015 IEEE 56th Annual Symposium on Foundations of Computer
  Science\/}, 1203--1217.

\bibitem[\protect\citeauthoryear{Deitsch}{Deitsch}{2018}]{elpv2018}
Deitsch, S. (2018).
\newblock A benchmark for visual identification of defective solar cells in
  electroluminescence imagery.
\newblock Available at \url{https://github.com/zae-bayern/elpv-dataset}.

\bibitem[\protect\citeauthoryear{Deitsch, Buerhop-Lutz, Sovetkin, Steland,
  Maier, Gallwitz, and Riess}{Deitsch et~al.}{2021}]{Deitsch2021}
Deitsch, S., C.~Buerhop-Lutz, E.~Sovetkin, A.~Steland, A.~Maier, F.~Gallwitz,
  and C.~Riess (2021).
\newblock Segmentation of photovoltaic module cells in uncalibrated
  electroluminescence images.
\newblock {\em Machine Vision and Applications\/}~{\em 32, DOI:
  10.1007/s00138-021-01191-9}.

\bibitem[\protect\citeauthoryear{Deitsch, Christlein, Berger, Buerhop-Lutz,
  Maier, Gallwitz, and Riess}{Deitsch et~al.}{2019}]{Deitsch2019}
Deitsch, S., V.~Christlein, S.~Berger, C.~Buerhop-Lutz, A.~Maier, F.~Gallwitz,
  and C.~Riess (2019).
\newblock Automatic classification of defective photovoltaic module cells in
  electroluminescence images.
\newblock {\em Solar Energy\/}~{\em 185}, 455--468.

\bibitem[\protect\citeauthoryear{Diakonikolas, Kane, and Stewart}{Diakonikolas
  et~al.}{2016}]{diakonikolas2016fourier}
Diakonikolas, I., D.~M. Kane, and A.~Stewart (2016).
\newblock The {Fourier} transform of {Poisson} multinomial distributions and
  its algorithmic applications.
\newblock In {\em Proceedings of the forty-eighth annual ACM symposium on
  Theory of Computing}, pp.\  1060--1073.

\bibitem[\protect\citeauthoryear{Dua and Graff}{Dua and Graff}{2017}]{Dua:2019}
Dua, D. and C.~Graff (2017).
\newblock {UCI} machine learning repository.
\newblock Available at
  \url{https://archive.ics.uci.edu/ml/datasets/AI4I+2020+Predictive+Maintenance+Dataset}.

\bibitem[\protect\citeauthoryear{Frigo and Johnson}{Frigo and
  Johnson}{2005}]{FrigoJohnson2005}
Frigo, M. and S.~Johnson (2005).
\newblock The design and implementation of fftw3.
\newblock {\em Proceedings of the IEEE\/}~{\em 93}, 216--231.

\bibitem[\protect\citeauthoryear{Goodfellow, Bengio, and Courville}{Goodfellow
  et~al.}{2016}]{Goodfellow-et-al-2016}
Goodfellow, I., Y.~Bengio, and A.~Courville (2016).
\newblock {\em Deep Learning}.
\newblock MIT Press.

\bibitem[\protect\citeauthoryear{Hong}{Hong}{2013}]{Hong2013}
Hong, Y. (2013).
\newblock On computing the distribution function for the {Poisson} binomial
  distribution.
\newblock {\em Computational Statistics and Data Analysis\/}~{\em 59}, 41--51.

\bibitem[\protect\citeauthoryear{Hong, Lin, Wang, and Junge}{Hong
  et~al.}{2022}]{R-PMD}
Hong, Y., Z.~Lin, Y.~Wang, and F.~Junge (2022).
\newblock {\em PoissonMultinomial: The Poisson-Multinomial Distribution}.
\newblock R package version 1.0.

\bibitem[\protect\citeauthoryear{Junge}{Junge}{2021}]{pbd}
Junge, F. (2021).
\newblock {\em {PoissonBinomial}: Efficient Computation of Ordinary and
  Generalized {Poisson} Binomial Distributions}.
\newblock R package version 1.2.4.

\bibitem[\protect\citeauthoryear{Schuessler}{Schuessler}{1999}]{Schuessler1999}
Schuessler, A.~A. (1999).
\newblock Ecological inference.
\newblock {\em Proceedings of the National Academy of Sciences\/}~{\em 96},
  10578--10581.

\bibitem[\protect\citeauthoryear{Zhang, Hong, and Balakrishnan}{Zhang
  et~al.}{2018}]{zhang2018generalized}
Zhang, M., Y.~Hong, and N.~Balakrishnan (2018).
\newblock The generalized {Poisson}-binomial distribution and the computation
  of its distribution function.
\newblock {\em Journal of Statistical Computation and Simulation\/}~{\em 88},
  1515--1527.

\end{thebibliography}
\end{document}